%% file: catalog.tex
\documentclass{emulateapj} 
\usepackage{amsmath,natbib,graphicx}
\usepackage{epsf}
\usepackage{epstopdf}
\usepackage{multirow}
\usepackage{epsfig}
\usepackage{color}
\usepackage{verbatim}
\DeclareGraphicsExtensions{.jpg,.pdf,.png,.eps,.ps}
\graphicspath{{FIGURES/}}

\DeclareGraphicsExtensions{.jpg,.pdf,.png,.eps,.ps}

\newcommand{\wCDM}{\mbox{wCDM}}

\newcommand{\mass}{\mbox{$M_{\mbox{\scriptsize 500}}$}}

\newcommand{\ltsima}{$\; \buildrel < \over \sim \;$}
\newcommand{\ltsim}{\lower.5ex\hbox{\ltsima}}
\newcommand{\sqdeg}{deg$^2$}

\newcommand{\muk}{\ensuremath{\mu {\rm K}}}
\newcommand{\mukcmb}{\ensuremath{\mu \mathrm{K}_\mathrm{CMB}}}
\newcommand{\nfalsefourf}{66}
\newcommand{\nfalsefive}{6}
\newcommand{\ncand}{224}
\newcommand{\nconfirmnew}{117}
\newcommand{\nconfirmnewspt}{135}
\newcommand{\nconfirmnewallspt}{144}
\newcommand{\nconfirmfourf}{158}

\newcommand{\ncosmo}{100}
\newcommand{\nfullsurvey}{550}
\newcommand{\chandra}{{\sl Chandra}}
\newcommand{\spitzer}{{\sl Spitzer}}
\newcommand{\hubble}{{\sl Hubble}}
\newcommand{\xmm}{{\sl XMM-Newton}}
\newcommand{\planck}{{\sl Planck}}
\newcommand{\rosat}{{\sl ROSAT}}
\newcommand{\snm}{$\zeta-M_{500}$}
\newcommand{\Yx}{\mbox{$Y_{X}$}}

\newcommand{\yxm}{\mbox{$Y_{X}-M_{500}$}}
\newcommand{\Tx}{\mbox{$T_{X}$}}
\newcommand{\Mg}{\mbox{$M_{g}$}}
\newcommand{\Txtheta}{\mbox{$T_{X}(r)$}}
\newcommand{\Mgr}{\mbox{$M_{g}(r)$}}
\newcommand{\sptcl}{\ensuremath{{\rm SPT_{CL}}}}
\newcommand{\wset}{\ensuremath{{\rm CMB + BAO + H_0 + SNe}}}
\newcommand{\nuset}{\ensuremath{{\rm CMB + BAO + H_0 }}}
\newcommand{\allset}{\ensuremath{{\rm CMB + BAO + H_0 + SNe + SPT_{CL}}}}

\newcommand{\medianzcosmo}{0.62}
\newcommand{\medianz}{0.55}
\newcommand{\medianm}{\ensuremath{3.3\times10^{14}\,M_\odot\,h_{70}^{-1}}}

\newcommand{\mcomplete}{\ensuremath{5\times10^{14}\,M_\odot\,h_{70}^{-1}}}
\newcommand{\mthreshold}{\ensuremath{2.5\times10^{14}\,M_\odot\,h_{70}^{-1}}}
\newcommand{\sigmaeightWCDM}{\ensuremath{0.807\pm0.027}}
\newcommand{\wWCDM}{\ensuremath{-1.010\pm0.058}}
\newcommand{\sigmaeightmnu}{\ensuremath{0.766\pm0.028}}
\newcommand{\mnu}{\ensuremath{0.38}}
\newcommand{\sigmaeightmnunoclus}{\ensuremath{0.775\pm0.041}}
\newcommand{\mnunoclus}{\ensuremath{0.44}}

\def \five {\textsc{ra5h30dec-55}}
\def \twthree {\textsc{ra23h30dec-55}}
\def \twonesix {\textsc{ra21hdec-60}}
\def \twonefif {\textsc{ra21hdec-50}}
\def \three {\textsc{ra3h30dec-60}}

\newcommand{\SN}{S/N}
\newcommand{\nSN}{\ensuremath{\xi}}

\newcommand{\fsz}{f_\mathrm{\mbox{\tiny{SZ}}}}

\newcommand{\lcdm}{\ensuremath{\Lambda{\rm CDM}}}

\newcommand{\be}{\begin{equation}}
\newcommand{\ee}{\end{equation}}
\newcommand{\bea}{\begin{eqnarray}}
\newcommand{\eea}{\end{eqnarray}}

\defcitealias{vanderlinde10}{V10}
\defcitealias{benson11}{B11}

\def\Berkeley{1}
\def\CfA{2}
\def\KICPChicago{3}
\def\PhysicsUChicago{4}
\def\CaseWestern{5}
\def\UChicago{6}
\def\Munich{7}
\def\MIT{8}
\def\NCSA{9}
\def\Harvard{10}
\def\ExcellenceCluster{11}
\def\EFIChicago{12}
\def\Miss{13}
\def\AAUChicago{14}
\def\ANL{15}
\def\NIST{16}
\def\PUC{17}
\def\McGill{18}
\def\UFlorida{19}
\def\Colorado{20}
\def\NASA{21}
\def\Davis{22}
\def\LBNL{23}
\def\Caltech{24}
\def\Arizona{25}
\def\Michigan{26}
\def\MPE{27}

\def\Minnesota{28}
\def\STScI{29}
\def\SAIC{30}
\def\Yale{31}
\def\BCCP{32}

\begin{document}

\title{Galaxy clusters discovered via the Sunyaev-Zel'dovich effect in the first 720 square degrees of the South Pole Telescope survey}

\author{ 
C.~L.~Reichardt\altaffilmark{\Berkeley},
B.~Stalder\altaffilmark{\CfA},
L.~E.~Bleem\altaffilmark{\KICPChicago,\PhysicsUChicago},
T.~E.~Montroy\altaffilmark{\CaseWestern},
K.~A.~Aird\altaffilmark{\UChicago},
K.~Andersson\altaffilmark{\Munich,\MIT},
R.~Armstrong\altaffilmark{\NCSA},
M.~L.~N.~Ashby\altaffilmark{\CfA},
M.~Bautz\altaffilmark{\MIT},
M.~Bayliss\altaffilmark{\Harvard}, 
G.~Bazin\altaffilmark{\Munich,\ExcellenceCluster},
B.~A.~Benson\altaffilmark{\KICPChicago,\EFIChicago},
M.~Brodwin\altaffilmark{\Miss},
J.~E.~Carlstrom\altaffilmark{\KICPChicago,\PhysicsUChicago,\EFIChicago,\AAUChicago,\ANL}, 
C.~L.~Chang\altaffilmark{\KICPChicago,\EFIChicago,\ANL}, 
H.~M. Cho\altaffilmark{\NIST}, 
A.~Clocchiatti\altaffilmark{\PUC},
T.~M.~Crawford\altaffilmark{\KICPChicago,\AAUChicago},
A.~T.~Crites\altaffilmark{\KICPChicago,\AAUChicago},
T.~de~Haan\altaffilmark{\McGill},
S.~Desai\altaffilmark{\Munich,\ExcellenceCluster},
M.~A.~Dobbs\altaffilmark{\McGill},
J.~P.~Dudley\altaffilmark{\McGill},
R.~J.~Foley\altaffilmark{\CfA}, 
W.~R.~Forman\altaffilmark{\CfA},
E.~M.~George\altaffilmark{\Berkeley},
M.~D.~Gladders\altaffilmark{\KICPChicago,\AAUChicago},
A.~H.~Gonzalez\altaffilmark{\UFlorida},
N.~W.~Halverson\altaffilmark{\Colorado},
N.~L.~Harrington\altaffilmark{\Berkeley},
F.~W.~High\altaffilmark{\KICPChicago,\AAUChicago}, 
G.~P.~Holder\altaffilmark{\McGill},
W.~L.~Holzapfel\altaffilmark{\Berkeley},
S.~Hoover\altaffilmark{\KICPChicago,\EFIChicago},
J.~D.~Hrubes\altaffilmark{\UChicago},
C.~Jones\altaffilmark{\CfA},
M.~Joy\altaffilmark{\NASA},
R.~Keisler\altaffilmark{\KICPChicago,\PhysicsUChicago},
L.~Knox\altaffilmark{\Davis},
A.~T.~Lee\altaffilmark{\Berkeley,\LBNL},
E.~M.~Leitch\altaffilmark{\KICPChicago,\AAUChicago},
J.~Liu\altaffilmark{\Munich,\ExcellenceCluster},
M.~Lueker\altaffilmark{\Berkeley,\Caltech},
D.~Luong-Van\altaffilmark{\UChicago},
A.~Mantz\altaffilmark{\KICPChicago},
D.~P.~Marrone\altaffilmark{\Arizona},
M.~McDonald\altaffilmark{\MIT},
J.~J.~McMahon\altaffilmark{\KICPChicago,\EFIChicago,\Michigan},
J.~Mehl\altaffilmark{\KICPChicago,\AAUChicago},
S.~S.~Meyer\altaffilmark{\KICPChicago,\PhysicsUChicago,\EFIChicago,\AAUChicago},
L.~Mocanu\altaffilmark{\KICPChicago,\AAUChicago},
J.~J.~Mohr\altaffilmark{\Munich,\ExcellenceCluster,\MPE},
S.~S.~Murray\altaffilmark{\CfA},
T.~Natoli,\altaffilmark{\KICPChicago,\PhysicsUChicago},
S.~Padin\altaffilmark{\KICPChicago,\AAUChicago,\Caltech},
T.~Plagge\altaffilmark{\KICPChicago,\AAUChicago},
C.~Pryke\altaffilmark{\Minnesota}, 
A.~Rest\altaffilmark{\STScI},
J.~Ruel\altaffilmark{\Harvard},
J.~E.~Ruhl\altaffilmark{\CaseWestern}, 
B.~R.~Saliwanchik\altaffilmark{\CaseWestern}, 
A.~Saro\altaffilmark{\Munich},
J.~T.~Sayre\altaffilmark{\CaseWestern}, 
K.~K.~Schaffer\altaffilmark{\KICPChicago,\EFIChicago,\SAIC}, 
L.~Shaw\altaffilmark{\McGill,\Yale},
E.~Shirokoff\altaffilmark{\Berkeley,\Caltech}, 
J.~Song\altaffilmark{\Michigan},
H.~G.~Spieler\altaffilmark{\LBNL},
Z.~Staniszewski\altaffilmark{\CaseWestern},
A.~A.~Stark\altaffilmark{\CfA}, 
K.~Story\altaffilmark{\KICPChicago,\PhysicsUChicago},
C.~W.~Stubbs\altaffilmark{\CfA,\Harvard}, 
R.~\v{S}uhada\altaffilmark{\Munich},
A.~van~Engelen\altaffilmark{\McGill},
K.~Vanderlinde\altaffilmark{\McGill},
J.~D.~Vieira\altaffilmark{\KICPChicago,\PhysicsUChicago,\Caltech},
A. Vikhlinin\altaffilmark{\CfA},
R.~Williamson\altaffilmark{\KICPChicago,\AAUChicago}, 
O.~Zahn\altaffilmark{\Berkeley,\BCCP},
and
A.~Zenteno\altaffilmark{\Munich,\ExcellenceCluster}
}

\altaffiltext{\Berkeley}{Department of Physics,
University of California, Berkeley, CA 94720}
\altaffiltext{\CfA}{Harvard-Smithsonian Center for Astrophysics,
60 Garden Street, Cambridge, MA 02138}
\altaffiltext{\KICPChicago}{Kavli Institute for Cosmological Physics,
University of Chicago,
5640 South Ellis Avenue, Chicago, IL 60637}
\altaffiltext{\PhysicsUChicago}{Department of Physics,
University of Chicago,
5640 South Ellis Avenue, Chicago, IL 60637}
\altaffiltext{\CaseWestern}{Physics Department, Center for Education and Research in Cosmology 
and Astrophysics, 
Case Western Reserve University,
Cleveland, OH 44106}
\altaffiltext{\UChicago}{University of Chicago,
5640 South Ellis Avenue, Chicago, IL 60637}
\altaffiltext{\Munich}{Department of Physics,
Ludwig-Maximilians-Universit\"{a}t,
Scheinerstr.\ 1, 81679 M\"{u}nchen, Germany}
\altaffiltext{\MIT}{MIT Kavli Institute for Astrophysics and Space
Research, Massachusetts Institute of Technology, 77 Massachusetts Avenue,
Cambridge, MA 02139}
\altaffiltext{\NCSA}{National Center for Supercomputing Applications,
University of Illinois, 1205 West Clark Street, Urbanan, IL 61801}
\altaffiltext{\Harvard}{Department of Physics, Harvard University, 17 Oxford Street, Cambridge, MA 02138}
\altaffiltext{\ExcellenceCluster}{Excellence Cluster Universe,
Boltzmannstr.\ 2, 85748 Garching, Germany}
\altaffiltext{\EFIChicago}{Enrico Fermi Institute,
University of Chicago,
5640 South Ellis Avenue, Chicago, IL 60637}
\altaffiltext{\Miss}{Department of Physics, University of Missouri, 5110 Rockhill Road, Kansas City, MO 64110}
\altaffiltext{\AAUChicago}{Department of Astronomy and Astrophysics,
University of Chicago,
5640 South Ellis Avenue, Chicago, IL 60637}
\altaffiltext{\ANL}{Argonne National Laboratory, 9700 S. Cass Avenue, Argonne, IL, USA 60439}
\altaffiltext{\NIST}{NIST Quantum Devices Group, 325 Broadway Mailcode 817.03, Boulder, CO, USA 80305}
\altaffiltext{\PUC}{Departamento de Astronom'a y Astrof'sica, PUC Casilla 306, Santiago 22, Chile}
\altaffiltext{\McGill}{Department of Physics,
McGill University,
3600 Rue University, Montreal, Quebec H3A 2T8, Canada}
\altaffiltext{\UFlorida}{Department of Astronomy, University of Florida, Gainesville, FL 32611}
\altaffiltext{\Colorado}{Department of Astrophysical and Planetary Sciences and Department of Physics,
University of Colorado,
Boulder, CO 80309}
\altaffiltext{\NASA}{Department of Space Science, VP62,
NASA Marshall Space Flight Center,
Huntsville, AL 35812}
\altaffiltext{\Davis}{Department of Physics, 
University of California, One Shields Avenue, Davis, CA 95616}
\altaffiltext{\LBNL}{Physics Division,
Lawrence Berkeley National Laboratory,
Berkeley, CA 94720}
\altaffiltext{\Caltech}{California Institute of Technology, 1200 E. California Blvd., Pasadena, CA 91125}
\altaffiltext{\Arizona}{Steward Observatory, University of Arizona, 933 North Cherry Avenue, Tucson, AZ 85721}
\altaffiltext{\Michigan}{Department of Physics, University of Michigan, 450 Church Street, Ann  
Arbor, MI, 48109}
\altaffiltext{\MPE}{Max-Planck-Institut f\"{u}r extraterrestrische Physik,
Giessenbachstr.\ 85748 Garching, Germany}
\altaffiltext{\Minnesota}{Physics Department, University of Minnesota, 116 Church Street S.E., Minneapolis, MN 55455}
\altaffiltext{\STScI}{Space Telescope Science Institute, 3700 San Martin
Dr., Baltimore, MD 21218}
\altaffiltext{\SAIC}{Liberal Arts Department, 
School of the Art Institute of Chicago, 
112 S Michigan Ave, Chicago, IL 60603}
\altaffiltext{\Yale}{Department of Physics, Yale University, P.O. Box 208210, New Haven,
CT 06520-8120}
\altaffiltext{\BCCP}{Berkeley Center for Cosmological Physics,
Department of Physics, University of California, and Lawrence Berkeley
National Labs, Berkeley, CA 94720}

\email{cr@bolo.berkeley.edu}

\begin{abstract}

We present a catalog of \ncand\ galaxy cluster candidates, selected through 
their Sunyaev-Zel'dovich (SZ) effect signature
in the first 720\,deg$^2$ of the South Pole Telescope (SPT) survey. 
This area was mapped with the SPT in the 2008 and 2009 austral winters to a depth of $\sim$\,$18\,\mukcmb$-arcmin at $150\,$GHz; 550\,deg$^2$ of it was also mapped to $\sim$\,$44\,\mukcmb$-arcmin at $95\,$GHz.  
Based on optical  imaging of all candidates and near-infrared imaging of the majority of candidates, we have found optical and/or infrared counterparts for \nconfirmfourf\ clusters.
Of these, \nconfirmnewspt{} were first identified as clusters in SPT data, including \nconfirmnew\ new discoveries reported in this work.
This catalog triples the number of confirmed galaxy clusters discovered through the SZ effect. 
We report photometrically derived (and in some cases spectroscopic) redshifts for confirmed clusters and redshift lower limits for the remaining candidates. 
The catalog extends to high redshift with a median redshift of $z = \medianz$ and maximum redshift of $z =  1.37$. 
Forty-five of the clusters have counterparts in the \rosat{} bright or faint source catalogs from which we estimate X-ray fluxes. 
Based on simulations, we expect the catalog to be nearly 100\% complete above $M_{500} \approx \mcomplete$ at $z \gtrsim 0.6$. 
There are 121 candidates detected at signal-to-noise greater than five, at which the catalog purity is measured to be 95\%. 
From this high-purity subsample, we exclude the $z < 0.3$ clusters and use the remaining 100 candidates to improve cosmological constraints following the method presented by \citet{benson11}.  
Adding the cluster data to CMB+BAO+H$_0$ data leads to a preference for non-zero neutrino masses while only slightly reducing the upper limit on the sum of neutrino masses to $\sum m_\nu < \mnu\,$eV (95\% CL). 
For  a spatially flat \wCDM \  cosmological model, 
the addition of this catalog to the CMB+BAO+H$_0$+SNe results yields $\sigma_8 =\sigmaeightWCDM$
and $w = \wWCDM$, improving the constraints on these parameters by a factor of 1.4 and 1.3, respectively. 
The larger cluster catalog presented in this work leads to slight
 improvements in cosmological constraints from those presented by
 \citet{benson11}. These cosmological constraints are currently limited by 
 uncertainty in the cluster mass calibration, not the size or quality of
 the cluster catalog.  A multi-wavelength observation program to improve the
 cluster mass calibration will make it possible to realize the full potential of the  
 final 2500\,deg$^2$ SPT cluster catalog to constrain cosmology.
\end{abstract}


\bigskip\bigskip

\section{Introduction}
\label{sec:intro}

\setcounter{footnote}{0}

Galaxy clusters are the largest collapsed objects in the Universe, and their abundance is exponentially sensitive to the growth of structure. 
Measurements of the abundance of galaxy clusters as a function of mass
and redshift have the potential to significantly improve current constraints on cosmological parameters,
including the equation of state of dark energy and the sum of the neutrino masses
\citep{wang98,haiman01,holder01b,battye03,molnar04,wang04,wang05,lima07,shimon11}.
To achieve this objective, a sample of galaxy clusters must have a well understood selection function, good mass estimates, and wide redshift extent. 

Most known galaxy clusters have been
identified by their optical properties 
or from their X-ray emission.
Clusters of galaxies contain anywhere from  tens 
to thousands of galaxies, but these galaxies account for a small
fraction of the total baryonic mass in a cluster (see, e.g, \citealt{allen11} for a review).  
Most of the baryons in  clusters are contained in the intra-cluster medium (ICM), the hot 
($10^7-10^8\,$K) X-ray-emitting plasma that pervades cluster environments.

\citet{sunyaev72} noted that such a plasma would also interact with
cosmic microwave background (CMB) photons via inverse Compton
scattering, causing a small spectral distortion of the CMB along the line
of sight to a cluster.  
This is called the  thermal Sunyaev Zel'dovich (SZ) effect.\footnote{In this work, `SZ effect' will refer to the thermal SZ effect unless specifically noted as the kinetic SZ effect.} 
The amplitude of the spectral distortion at a given position on the
sky is proportional to the integrated electron
pressure along the line of sight. 
Therefore, the integrated thermal SZ (tSZ) flux
is a direct measure of the total thermal energy of the ICM, and the
SZ flux is thus expected to be a robust proxy for total cluster mass
\citep{barbosa96, holder01a, motl05}.
Additionally, the SZ surface brightness is independent of redshift.  
As a result, SZ surveys with sufficient angular resolution have the 
potential to deliver nearly mass-limited cluster samples
over a wide redshift range \citep{carlstrom02}.
Such a cluster sample can provide a growth-based test of dark energy
to complement the distance-based tests provided by supernovae
\citep[e.g.,][]{riess98,perlmutter99a}; it can also probe the sum of the neutrino masses. 
Recent results \citep[e.g.,][]{vikhlinin09, mantz10b,benson11} have demonstrated the
power of such tests to constrain cosmological models and parameters.

However, the SZ signal is faint, exceeding a few hundred \muk{}  for only the most massive (and rare) galaxy clusters. 
As a result,  experiments have only recently achieved the requisite sensitivity to discover previously unknown galaxy clusters. 
Since the first discovery of clusters using South Pole Telescope (SPT) data \citep{staniszewski09}, SZ-selected galaxy cluster catalogs have been produced by the SPT, Atacama Cosmology Telescope (ACT), and \planck\ collaborations \citep{vanderlinde10,williamson11,marriage11b,planck11-5.1a}. 
In total, roughly 40 previously unknown clusters discovered via the SZ effect have been 
published to date.

This is the third SPT cluster catalog and fourth SPT cosmological analysis based on galaxy cluster counts. 
\citet[hereafter V10]{vanderlinde10} presented the first SZ-selected catalog, consisting of 
21 optically confirmed galaxy clusters found in 2008 SPT data.
\citetalias{vanderlinde10} also investigated the cosmological implications of these clusters,
using a simulation-calibrated mass scaling relation.
The second SPT cluster catalog and cosmological analysis \citep[hereafter W11]{williamson11} used the most massive galaxy clusters discovered in the entire 2500\,deg$^2$ SPT survey region to test for non-Gaussianity and consistency with \lcdm\ . 
In the third analysis, \citet[hereafter B11]{benson11} 
 developed a method to combine X-ray  data with the SZ observations, and thereby improve the cluster mass estimates.
B11 used this method to improve the cosmological constraints from the \citetalias{vanderlinde10} cluster sample. 

In this work, we present a catalog of \ncand\ SZ-identified galaxy cluster candidates above $4.5\,\sigma$ from the first 720\,deg$^2$ of the SPT survey.
Using follow-up optical imaging of all candidates and near-infrared (NIR) imaging for a subset,  we estimate redshifts for \nconfirmfourf\ of the candidates and
 calculate lower redshift limits for the remaining candidates, which are either too distant to identify with current optical/NIR observations or are spurious detections in the SPT data. The details of the optical and NIR data and redshift estimates are given in a companion paper (J.~Song et al. in prep., hereafter S12)\nocite{song12b}. Here we summarize the observations and report the resulting redshifts.
The clusters with clear optical/NIR counterparts include \nconfirmnew\ new discoveries, which increases the number of clusters discovered with the SPT to \nconfirmnewallspt{} and triples the total number of SZ-identified clusters.  
Simulations are used to characterize the SPT cluster selection function. 
We combine the cluster list with the improved mass-scaling relation from B11 to improve cosmological constraints on large-scale structure, neutrino masses, and the dark energy equation of state.

The paper is organized as follows.
We describe the observations and map-making in \S\ref{sec:obs}. 
The extraction of galaxy clusters from the maps is detailed in \S\ref{sec:extract}. 
The optical followup campaign and the resulting redshifts are presented in \S\ref{sec:optical}. 
In \S\ref{sec:catalog}, we present the complete catalog of galaxy cluster candidates. 
We review the B11 method for simultaneously constraining cosmological and 
scaling relation parameters in \S\ref{sec:cosmo}, and we 
discuss the cosmological constraints from this cluster catalog and prospects for future improvement in 
\S\ref{sec:constraints} before concluding in \S\ref{sec:conclusions}.

\section{Observations and Data Reduction}
\label{sec:obs}

\subsection{Telescope and Observations}
\label{subsec:telobs}

The South Pole Telescope (SPT) is a 10-meter telescope designed to
survey a large area of the sky at millimeter wavelengths with 
arcminute angular resolution \citep{ruhl04, padin08,
carlstrom11}.  The first SPT receiver was a three-band (95, 150, and 220~GHz)
bolometer camera optimized for studying the primary CMB anisotropy and the tSZ effect. 
From the time the SPT was commissioned through the end of 2011, the majority of 
observing time was spent on the recently completed $2500\,{\rm deg}^2$ SPT survey. 
The cluster catalog presented in this paper is derived from the first $720\,{\rm deg}^2$ of this survey.  
This area was observed during the Austral winters of 2008 and 2009. 
In addition to the early SPT galaxy cluster results discussed in \S\ref{sec:intro}, science results from early
subsets of the survey data 
have included measurements of the primary and secondary CMB anisotropy \citep{keisler11,lueker10, shirokoff11,reichardt11}, a measurement of gravitational lensing of the CMB \citep{vanengelen12}, and the discovery of a new population of extremely bright submillimeter galaxies \citep{vieira10}. 

For cluster-finding, we use data from the SPT 95\,GHz and 150\,GHz frequency bands. 
The effective bandcenters for a non-relativistic tSZ spectrum are 97.6\,GHz and 152.9\,GHz. 
The $220\,$GHz band is centered near the tSZ null, so it contains effectively no SZ cluster signal. 
In the 2008 observing season, the 480 detectors at 150\,GHz performed well, but the 
95\,GHz detectors did not meet specifications.
The receiver was reconfigured  for the 2009 observing season with   
640  detectors at $150\,$GHz and 160 new
detectors at $95\,$GHz. 
We observed roughly $170\,{\rm deg}^2$ in two fields in 2008 and 550\,\sqdeg\ in three fields in 2009. 
Each field was observed to a minimum depth of $18\,\mukcmb$-arcmin at $150\,$GHz.\footnote{Throughout this work, the unit $K_\mathrm{CMB}$ refers to equivalent fluctuations in the CMB temperature, i.e.,~the temperature fluctuation of a 2.73$\,$K blackbody that would be required to produce the same power fluctuation.} 
The 2009 fields were observed to a minimum depth of $44\,\mukcmb$-arcmin at $95\,$GHz.
The SPT map of the first of the two 2008 fields is publicly available \citep{schaffer11}.

The standard operating mode of the SPT is to observe a target field by scanning back and forth in azimuth across the field followed by a step in elevation \citep{schaffer11}.  
One field (\twonefif) was observed with a hybrid scan strategy including  scans at both constant elevation and constant azimuth. 
This scan strategy changes the filtered point spread function for this field compared to the rest of the data, which affects the SPT signal-to-noise to cluster mass scaling relations presented in \S\ref{sec:szscale}.

The SPT beams have been measured using a combination of bright active galactic nuclei (AGN) in the survey fields and 
targeted observations of planets 
\citep{shirokoff11, keisler11}.
The SPT beam can be described by  a main lobe and a diffuse sidelobe. 
For compact sources such as galaxy clusters, the effect of the sidelobe is degenerate with a calibration factor, and we choose to fold it into the calibration.
The SPT main lobe beam is well-described by a Gaussian with FWHM = $1\farcm6$ and $1\farcm19$ at 95 and 150 GHz respectively. 
The 2009 data in this work are calibrated using observations of RCW38, a galactic HII region \citep[][W11]{staniszewski09}, while the 2008 data are calibrated by cross-correlating 
dedicated SPT observations of large patches of sky with WMAP observations of those same 
regions (V10).

The pointing model is determined using daily observations of galactic HII regions and sensors on the telescope structure sensitive to temperature and mechanical movement \citep{schaffer11}. 
The final pointing in the maps is checked against the positions of radio sources in the
Australia Telescope 20 GHz survey \citep[AT20G,][]{murphy10}, which has 
positional accuracy to better than 1 arcsec.
The absolute SPT pointing measured in this way is accurate to 3 arcsec. 
The RMS pointing uncertainty in the maps is 7 arcsec.

\subsection{Map Making}

The map-making algorithm for the SPT data has been described in detail in \citet{lueker10}, \citet{shirokoff11}, and \citetalias{vanderlinde10}.
In overview, the first step is to apply a relative calibration to the time-ordered data (TOD) and then band-pass filter the TOD.
Correlated atmospheric signals are removed by subtracting the mean signal across a set of adjacent bolometers.
We mask bright point sources detected at $>$$\,5\,\sigma$ at $150\,$GHz ($>$$\,\sim 6\,$mJy) before filtering.
The pointing for each detector is reconstructed, and the data from each detector are coadded into a map with inverse-noise weighting.

The maps (and cluster list) for the 2008 season are identical to those presented by \citetalias{vanderlinde10}.
Maps for the 2009 season have several small differences in the filtering detailed below:

\begin{itemize}
\item

In \citetalias{vanderlinde10}, the bandpass filter was set by a high-pass filter (HPF) at 0.25 Hz and a low-pass filter at 25 Hz. 
In 2009, different fields were observed at different scan speeds, so we choose to define the HPF with respect to angular multipole $\ell$. 
The HPF of the 2009 data is at $\ell=400$; the \citetalias{vanderlinde10} HPF corresponds to $\ell\simeq350$. 
As in \citetalias{vanderlinde10}, the HPF is implemented by removing a set of sines and cosines from each scan across the field.
We supplement the Fourier mode removal by first fitting and removing a 9$^{\rm th}$ order Legendre polynomial from each scan. 
The higher order (\citetalias{vanderlinde10} used first order) is necessitated by the large atmospheric modulation introduced by the subset of observations which scan in elevation.
Depending on the observation, this filter acts as a high-pass filter in either the R.A.~or decl.~direction.

\item

\citetalias{vanderlinde10} removed both the mean and slope  across the two-dimensional array of all detectors at a single frequency. 
The 2009 data have four times as many $150\,$GHz detectors as $95\,$GHz detectors so the \citetalias{vanderlinde10} scheme would result in different common mode removal at each frequency.
Instead, we follow the treatment in \citet{shirokoff11} and remove the mean across sets of neighboring detectors. 
The $150\,$GHz detectors are divided into four sets based on their position in the focal plane and the $95\,$GHz detectors are treated as a single set.
This filter  set choice produces nearly identical filtering at 95 and $150\,$GHz.

\end{itemize}

\section{Cluster Extraction}
\label{sec:extract}

The procedure used in this work to identify  SZ  galaxy cluster candidates is
identical to that used by W11.  We summarize the procedure here and refer the 
reader to W11 for more details.

Most of the SPT fields have been observed in three frequency bands, centered at 95, 150, and 220\,GHz. 
(Roughly one quarter of the sky area considered in this work was observed in 2008 without 95\,GHz coverage.) 
Each map at a given observing frequency contains contributions from
multiple astrophysical signals, and each signal has its own spatial and
spectral properties.  
Because the maps are calibrated in CMB fluctuation temperature units, 
primary CMB fluctuations and the (small) signal from the kinetic SZ (kSZ) effect contribute equally to all frequencies. 
Emissive radio galaxies appear in all frequencies with a falling spectral index, while dusty, star-forming galaxies appear with a rising spectral index. 
Most notably, the $95$~GHz and $150$~GHz maps contain 
the tSZ effect signal from galaxy clusters.
Because the spectral signature of the tSZ effect is known (up to a small relativistic
correction), and because we can roughly predict the spatial profile of the tSZ signal
from galaxy clusters, we can combine the maps from the different bands, weighted in
spatial frequency space by the expected cluster profile, to maximize the signal-to-noise 
of the tSZ effect from clusters.

Under certain assumptions about the noise, the astrophysical 
contaminants, and the source profile, it can be shown
\citep[e.g.,][]{melin06} that the optimal way
to extract a cluster-shaped tSZ signal from our data is to construct a simultaneous
spatial-spectral filter, given by
\begin{equation}
\boldsymbol{\psi}(k_x,k_y,\nu_i) = \sigma_\psi^{-2} \ 
\sum_j \mathbf{N}_{ij}^{-1}(k_x,k_y) \fsz(\nu_j) S_\mathrm{filt}(k_x,k_y,\nu_j).
\label{eqn:optfilt}
\end{equation}
Here, $\sigma_\psi^{-2}$ is the predicted variance in the filtered map
\begin{align}
\sigma_\psi^{-2} = 
\sum_{i,j}& \fsz(\nu_i) S_\mathrm{filt}(k_x,k_y,\nu_i) \ \mathbf{N}_{ij}^{-1}(k_x,k_y) \ \times \\
\nonumber & \fsz(\nu_j) S_\mathrm{filt}(k_x,k_y,\nu_j),
\end{align}
$S_\mathrm{filt}$ is the assumed cluster profile convolved with the instrument beam 
and any filtering performed in the mapmapking step, $\mathbf{N}_{ij}$ is the band-band
noise covariance matrix (including contributions from astrophysical signals other than
cluster tSZ), and $\fsz$ encodes the frequency scaling of the tSZ effect relative to primary 
CMB fluctuations \citep[e.g.,][]{carlstrom02}.

As in W11, our model for the astrophysical contribution to $\mathbf{N}_{ij}$
is a combination of primary and lensed CMB fluctuations, point sources below the SPT
detection threshold, kSZ, and tSZ from clusters below the SPT detection threshold. 
The assumptions about the spatial and spectral shapes of each component are identical
to those in W11.  As in all previous SPT cluster survey publications, the assumed cluster profile  
 is described by a projected spherical isothermal $\beta$-model \citep{cavaliere76}, with $\beta$ fixed to 1. 
Twelve different matched filters were constructed and applied to the data, 
each with a different core radius, spaced evenly between $0.25^\prime$ and $3.0^\prime$.
As in previous work, point sources detected above $5\, \sigma$ at 150\,GHz were 
masked out to a radius of $4^\prime$,  with the value inside that radius set to the average 
of the surrounding pixels from $4^\prime < r < 4.5^\prime$.
Furthermore,  cluster detections within $8^\prime$ of one of these $>$\,$5\,\sigma$ point sources were rejected.
Clusters were extracted from the filtered maps with the process used in all
previous SPT cluster work and described by \citetalias{vanderlinde10}.  As in \citetalias{vanderlinde10} and W11, we refer to the 
detection significance maximized across all twelve 
matched filters as $\nSN$, and we use $\nSN$ as the primary SZ observable.
As in W11, we use only $95$ (where available) and $150$~GHz data to extract clusters, as adding 
the $220$~GHz data does not result in measurable improvement in cluster yield
(see W11 for details).

\subsection{Simulations}
\label{sec:sims}

We use simulations to determine priors on the SZ scaling relations discussed in \S\ref{sec:szscale} as well as the expected false detection rate for the sample. 
Simulated sky realizations are filtered to match the real data, and noise realizations based on the measured map noise properties are added. 

Each simulated sky is a Gaussian realization of the sum of the  best-fit lensed WMAP7 $\Lambda$CDM primary CMB model, a kSZ model, and point source contributions. 
The kSZ power spectrum is taken from the \citet{sehgal10} simulations and has an amplitude, $D_l = l(l+1)C_l$, of $2.05\, \mu{\rm K}^2$ at $\ell=3000$. 
We include both Poisson and clustered point sources. 
The Poisson contribution reflects both radio source and dusty, star-forming galaxy (DSFG) populations.
 The amplitude of the radio source term is set by the \citet{dezotti05} model source counts to an amplitude $D_{3000}^{r} = 1.28\, \mu{\rm K}^2$ at $150\,$GHz with an assumed spectral index of $\alpha_r=-0.6$ (defined by flux $\propto \nu^\alpha$). 
 The amplitude of the Poisson DSFG term at 150\,GHz is $D_{3000}^{p} = 7.7 \,\mu{\rm K}^2$. 
 Finally, the clustered DSFG component is modeled by a $D_\ell \propto \ell$ term normalized to $D_{3000}^{c} = 5.9 \,\mu{\rm K}^2$ at 150\,GHz. 
 The DSFG terms have an assumed spectral index of 3.6.
The amplitude of each component was selected to be consistent with the \citet{shirokoff11} bandpowers.

For the determination of the SZ detection significance to cluster
 mass scaling, we also add a map of the tSZ effect; this tSZ map is not included when estimating the false detection rate. 
The tSZ map is drawn from a 4000\,\sqdeg\ simulation by \citet{shaw10}. 
Note that the limited sky area in this simulation means that we reuse the same tSZ maps between different fields in order to get 100 realizations.  
This limitation does not exist for the Gaussian realizations.

\subsection{Expected false detection rates}
\label{subsec:simpurity}
\begin{figure}[]
\centering
\includegraphics[width=.45\textwidth]{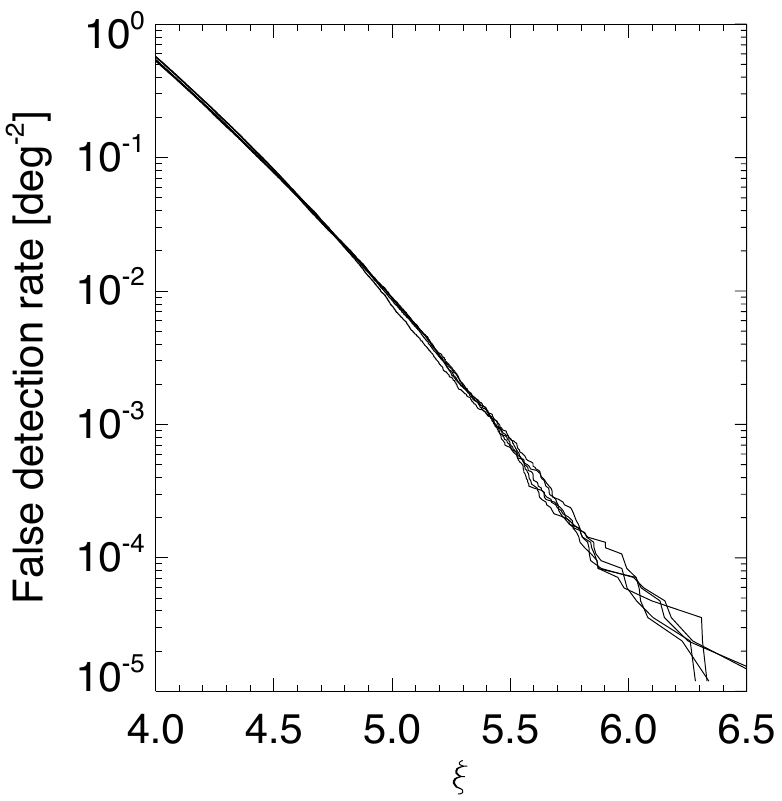}
  \caption[]{Simulated cumulative false detection rates, for each of the five fields,
    as a function of lower $\SN$ threshold (\nSN).
   No significant differences between the fields are observed. 
    The vertical axis shows the number density of false detections above a given $\SN$.
    \\ }
\label{fig:sz_falserate}
\end{figure}

We use the simulations described above, omitting the tSZ component, in order to estimate the rate of false
detections arising from noise and non-cluster astrophysical signals. 
The resulting rates are shown in Fig.~\ref{fig:sz_falserate}. 
As expected, the false detection rate is essentially indistinguishable between the fields;  
there are the same number of $N\sigma$ noise fluctuations per unit area. 
The simulations lead to a prediction of 6.4 false detections in the $>$$\,5\,\sigma$ catalog and 59 false detections in the  $>$$\,4.5\,\sigma$ catalog.

\subsection{Integrated Comptonization}
\label{subsec:yint}

For each cluster candidate, we estimate the integrated Comptonization by fitting the cluster to a projected spherical
$\beta$-model with $\beta=1$
\begin{equation}
Y(\theta) = y_0 \left (1+\frac{\theta^2}{\theta_c^2} \right )^{-1},
\end{equation}
where $y_0$ is peak Comptonization and $\theta_c$ is the angular
radius of the cluster core. The integrated Comptonization is defined as
\begin{equation}\label{eqn:yint}
Y_{\theta_I} = 2\pi \int_0^{\theta_I} Y(\theta) d\theta.
\end{equation}
In Table \ref{tab:catalog}, we set $\theta_I =
1^{\prime}$ and report $Y_{1^{\prime}}$.
 We expect measurements of $Y_{1^{\prime}}$ to be robust despite the
well known degeneracy between $\theta_c$ and the central Compton
parameter $y_0$ for observations that do not resolve the cluster core
\citep[e.g.,][]{planck11-5.1a}.

The likelihood of a set of cluster model parameters
$\mathcal{H}$ given our set of observed maps $D_{\nu}(\bar{x})$ is
defined as 
\begin{eqnarray}
\log(&P&(D|\mathcal{H})) = \\
\nonumber && -\frac{1}{2}\sum_{\bar{k},\nu1,\nu2}\frac{(\widetilde{D}_{\nu_1}(\bar{k})-\widetilde{s}^{\mathcal{H}}_{\nu_1}(\bar{k}))(\widetilde{D}_{\nu_2}(\bar{k})-\widetilde{s}^{\mathcal{H}}_{\nu_2}(\bar{k}))^*}{N_{\nu_1
    \nu_2}(\bar{k})},
\end{eqnarray}
where $\widetilde{D}_{\nu}(\bar{k})$ is the Fourier transform of the
map for frequency $\nu$, $\widetilde{s}^{\mathcal{H}}_{\nu}$ is the
frequency-dependent Fourier transform of the cluster model for
parameters set $\mathcal{H}$ which we define as ($\bar{x},~\theta_c,~y_0$), and $N_{\nu_1\nu_2}(\bar{k})$ is the
frequency-dependent covariance matrix of the set of maps which
accounts for the same noise and astrophysical components used in 
the matched filter analysis. For the cluster profile, we use the
projected spherical $\beta$-model defined above. 
We only fit the profile within $\theta < 5 \theta_c$.

We use the Rapid Gridded Likelihood Evaluation
(RGLE) method (T.~Montroy et al., in prep) \nocite{montroy12} to evaluate the cluster likelihood and compute $Y_{1^{\prime}}$.
The RGLE method is based on computing the likelihood for each cluster candidate on a fixed grid in parameter space. 
In this case, it is a four-dimensional grid over the parameters set $\mathcal{H}$. 
We define the extent of the grid as follows. 
The 2D position, $\bar{x}$, is constrained to be within $1\farcm5$ of the matched filter position. 
The central decrement is allowed to range from $-4.3 \times 10^{-4}$ to
$2.2 \times 10^{-3}$; this prior does not impact the results. 
The core radius, $\theta_c$, is required to be between $0^{\prime}$ and $7\farcm 5$.
For cluster candidates at $z > 0.125$, we additionally limit the physical core radius ($r_c$) of the cluster to be less than
1~Mpc. 
We translate between $r_c$ and $\theta_c$ based on the redshift of each cluster candidate (or redshift lower limit if unconfirmed).
A core radius of 1~Mpc is much larger than the typical cluster size,  so this limit allows
full exploration of the likelihood degeneracy between $Y_0$ and
$\theta_c$ while reducing the chance of bias due to noise fluctuations
on scales much larger than the expected cluster size.

 To compute the probability distribution for $Y_{1^{\prime}}$, we first marginalize the four-dimensional grid over position (i.e., $\bar{x}$) to determine the two-dimensional likelihood surface for $(\theta_c, y_0)$. 
 The value of $Y_{1^{\prime}}$ at each $(\theta_c, Y_0)$ is calculated from Eqn.~\ref{eqn:yint} with $\theta_I = 1^{\prime}$.
 Formally, the likelihood for a given value of $Y_{1^{\prime}}$ can be computed by integrating the likelihood surface over curves of constant 
 $Y_{1^{\prime}}$, 
\begin{equation}
P(D|Y_{1^{\prime}} = Y_i) = \int dY_0 d \theta_c P(D|Y_0,\theta_c) \delta(Y_{1^{\prime}}(Y_0,\theta_c)-Y_i).
\end{equation}
The median value and $68\%$ confidence intervals for  $Y_{1^{\prime}}$ are determined from this likelihood function.

When applying the RGLE method to the SPT maps in order to estimate $Y_{1^{\prime}}$, we use the calibration and beam shapes reported in \cite{reichardt11}.  We note that
for the 2009 data, these are slightly different from the calibration and beam model described in \S\ref{subsec:telobs} and used in cluster finding in this work. 
We use maps at $95$\,GHz (where available) and $150$\,GHz to estimate the cluster
properties. To limit contamination from point sources, we use maps where
previously identified point sources have been subtracted. The point
source amplitudes are estimated using a variant of the RGLE which fits
for the point source amplitudes given the beam shape. 
The point source subtraction significantly changes $Y_{1^{\prime}}$ for very few clusters since all affected point sources are at least 8$^\prime$ away from any cluster candidate.

The RGLE method was previously used in \citet{story11} to compute
integrated Comptonization for SPT follow-up observations of
\planck\ ESZ cluster candidates \citep{planck11-5.1a}. 
The method has been verified by extensive simulations; 
we have also checked that the RGLE method produces comparable results to an alternative method based on Markov Chain Monte Carlo-based sampling of the likelihood surface (B. Saliwanchik et al., in prep).\nocite{saliwanchik12}

\section{External data}
\label{sec:optical}

In this section, we briefly describe the optical, NIR, 
and X-ray data associated with this catalog. 
The optical/NIR followup strategy and analysis methods are summarized here and discussed 
in detail by S12 \nocite{song12b}.
We also summarize the dedicated X-ray measurements of 14 SPT clusters, measurements
which are used in the cosmological analysis here and
which have been discussed in detail in previous SPT publications.
Finally, we report X-ray fluxes and luminosities for all candidates 
that have identified counterparts in the {\it Roentgensatellit} (\rosat) all-sky survey. 

\subsection{Optical and NIR Data}

Every SPT-selected cluster candidate is followed up with optical imaging observations,
and many candidates are also targeted with NIR imaging.  Our strategy has evolved over time in order to utilize limited telescope resources to measure redshifts for the majority of cluster candidates. 
 Briefly, the SPT candidates are pre-screened with Digitized Sky Survey (DSS) data.  Candidates that appear to be at low redshift are followed up with the $1-$meter Swope telescope.  Candidates 
that appear to be at high redshift (i.e., that do not appear in DSS images) are targeted 
 with the 4-m Blanco telescope at CTIO or the 6.5-m Magellan
telescopes at Las Campanas Observatory.
The  $4-6$~meter class observing is performed using an adaptive strategy, wherein candidates 
are imaged for a short time in three bands, then with a second pass in two bands if the cluster has not been detected.  The second-pass imaging is designed to reach depths sufficient to confirm a $z\sim0.9$ cluster.  Given weather and other constraints, not all candidates were observed to full depth.

Space-based NIR observations with \spitzer/IRAC were obtained at 3.6\,$\mu$m and 4.5\,$\mu$m for the subset of candidates
above a threshold of $\xi=4.8$  ($\xi \ge 4.5$ for 350\,deg$^2$ of SPT coverage) that were not identified as low redshift clusters in DSS data. 
Candidates that were not imaged with \spitzer{} ---  and for which redshifts could not be estimated
from the acquired optical data --- were targeted with $K_s$-band observations with the NEWFIRM camera on the Blanco 4-m.

A number of clusters were also observed using either long-slit or
multi-slit spectrographs in subsequent follow-up projects.
A robust biweight location estimator \citep{beers90} is used to determine the
cluster spectroscopic redshifts from ensemble spectra of member galaxies.
Of the clusters in this work, 57 have spectroscopic redshifts, either from 
the literature or from our targeted observations.  The redshifts are shown
in Table \ref{tab:catalog}, and the source for every spectroscopic redshift
is presented by S12.

\subsection{Optical/NIR Imaging Data Reduction and Redshift Determination}

All optical images are processed using the PHOTPIPE analysis pipeline
 \citep{rest05a,miknaitis07}, as was done in previous SPT optical follow up
analyses \citep{high10,williamson11,story11}.
A separate reduction of the optical data from the Blanco Mosaic-II imager 
is performed using a version 
of the Dark Energy Survey (DES) data management pipeline \citep{mohr08,desai11}, which will eventually be used for analysis of data once the DES begins.
The \spitzer{}/IRAC imaging data are processed from the standard online pipeline system and analyzed as described in \citet{ashby09}; NEWFIRM data are reduced using the FATBOY pipeline \citep{eikenberry06}.

Redshifts are estimated for each candidate using three 
methods as described by S12\nocite{song12b}.  The first two methods are based on the identification of 
red-sequence overdensities and are described in detail in \citet{high10}
and \citet{song12a}, respectively.  The third method estimates photometric
redshifts for individual galaxies using the ANNz algorithm \citep{collister04}, 
and cluster redshifts are estimated by measuring a peak in a manually-selected red
galaxy photometric redshift distribution.  For a given cluster candidate, redshift 
estimates from the three methods are compared, outliers are flagged,
and a combined redshift estimate is produced.  
In cases where only the Spitzer/IRAC 3.6\,$\mu$m  and 4.5\,$\mu$m  data are deep enough to detect the cluster, 
we use the \citet{high10} method to estimate the redshift.  
Tests confirm this to be reliable at $z>0.7$ and a similar method is described in \citet{stern05} and \citet{papovich08}.
These redshifts and associated uncertainties are shown in Table \ref{tab:catalog}.
If none of the three methods is successful at estimating a redshift for a given
candidate, we report a lower redshift limit based upon the depth of the follow-up imaging.

\begin{figure*}[]
\centering
\plottwo{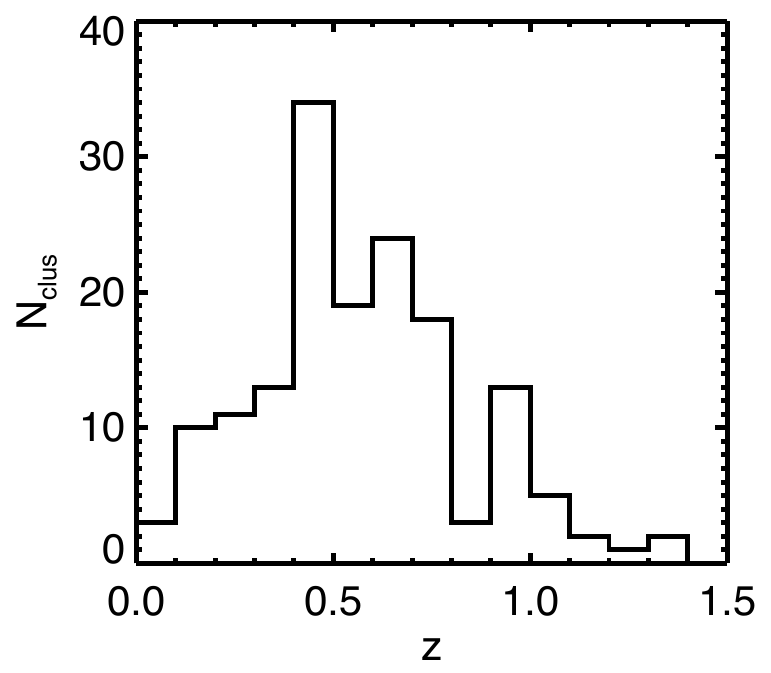}{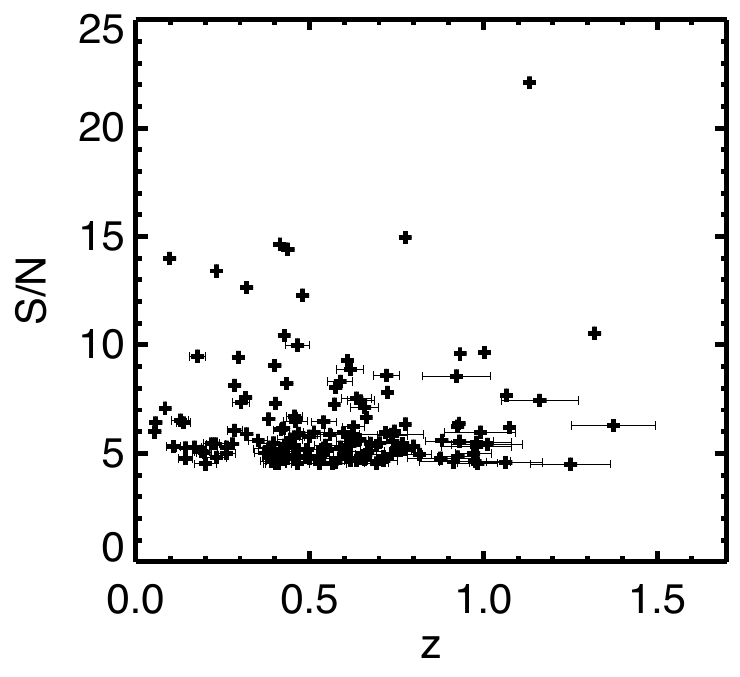}
  \caption[]{\textbf{\textit{Left panel:}} Redshift histogram for the optically confirmed, $\nSN > 4.5$ galaxy clusters in this sample. 
  The median redshift of the sample is \medianz. 
  The median redshift of the sample used in the cosmological analysis ($z > 0.3$ and $\nSN > 5$) is \medianzcosmo. 
  \textbf{\textit{Right panel:}} 
  Detection significance versus redshift for all optically confirmed galaxy clusters in this sample. 
    \\ }
\label{fig:zhist}
\end{figure*}

\begin{figure}[]
\centering
\includegraphics[width=.45\textwidth]{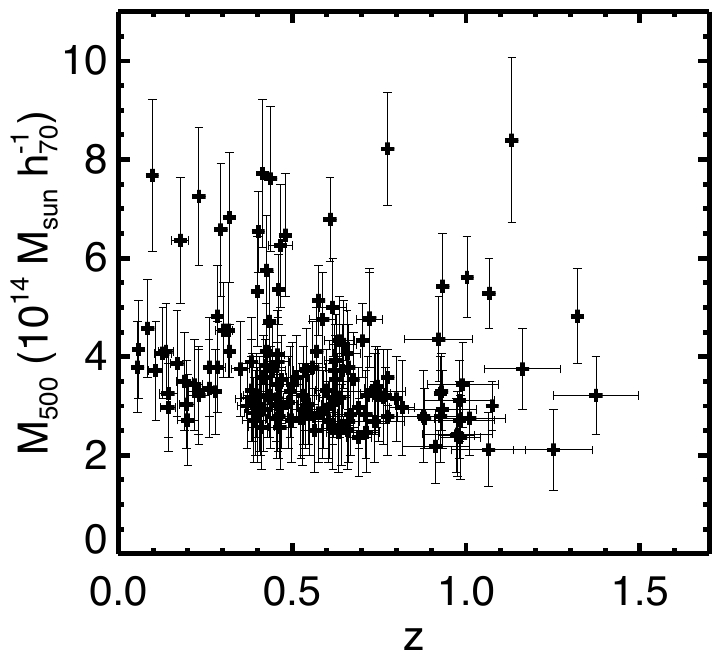}
  \caption[]{Cluster mass estimates versus redshift for all optically confirmed galaxy clusters in this sample. 
  The reported mass has been deboosted and marginalized over the allowed set of cosmological and scaling relation parameters for a \lcdm{} cosmology. 
    \\ }
\label{fig:mass}
\end{figure}

\begin{figure*}[]
\centering
\includegraphics[width=.9\textwidth]{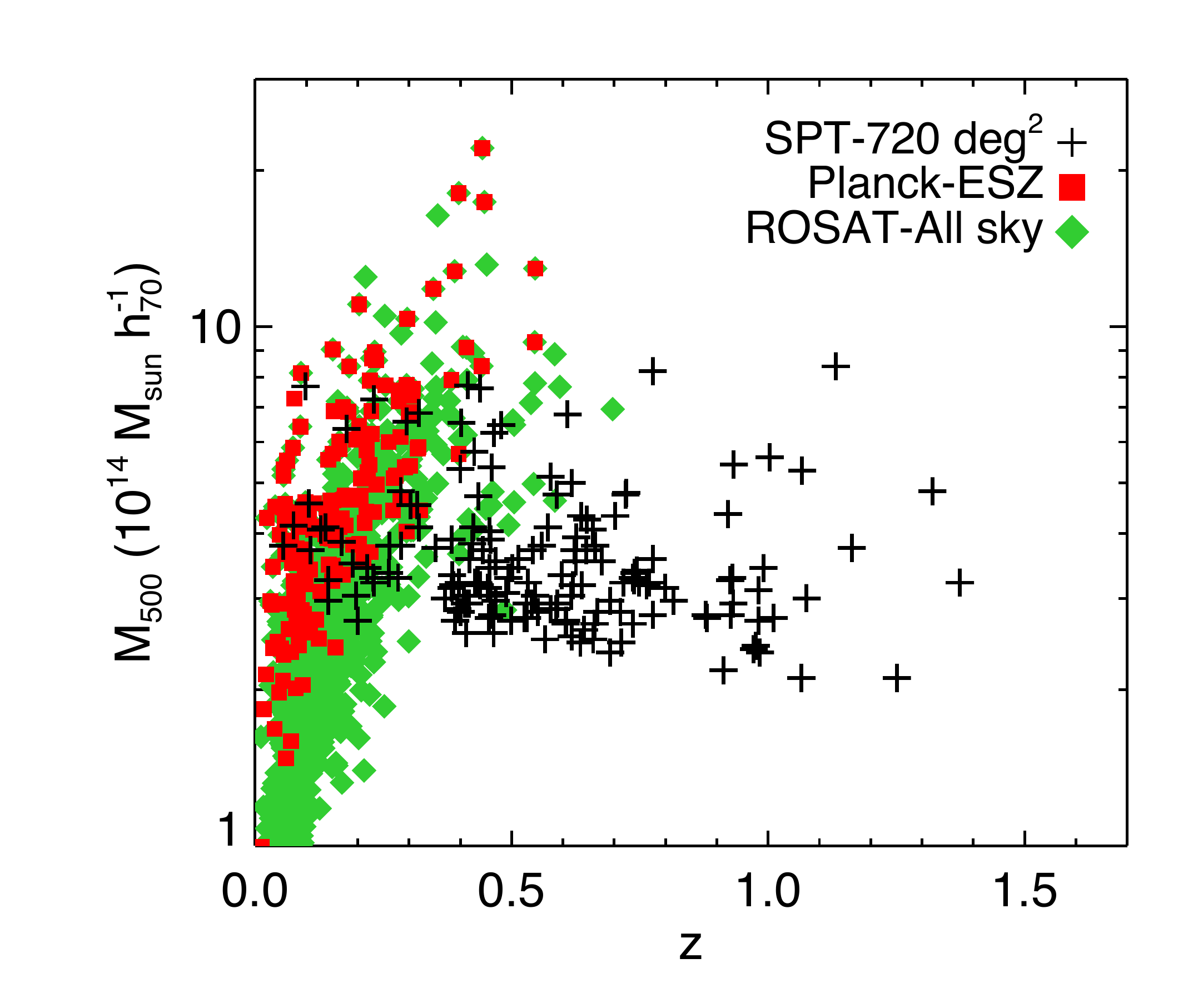}
  \caption[]{ Mass estimates versus redshift for three cluster samples: (1) optically-confirmed SZ-selected galaxy clusters from the SPT survey, (2) SZ-selected galaxy clusters from the \planck{} survey \citep{planck11-5.1a}, and (3) X-ray selected galaxy clusters from the \rosat{} all-sky survey \citep{piffaretti11}.  
High resolution SZ surveys, such as that performed with the SPT, uniquely have a  nearly redshift independent selection function. 
  The redshift dependent selection in the \planck{} survey is due to beam dilution; the redshift dependence of the \rosat{} catalog is due to cosmological dimming. 
    \\ }
\label{fig:masscomp}
\end{figure*}

\subsection{X-ray data}

\subsubsection{Dedicated X-ray Observations of SPT Clusters} \label{sec:dedicatedxray}

As first reported in \citet[][A11]{andersson11}, we have obtained \chandra \ and \xmm \ data 
on 15 of the highest \SN\ clusters from the 2008 SPT survey fields, including 14 clusters in 
the redshift range used in the cosmological analysis in this work ($z > 0.3$).  
B11 updated the X-ray observables for some clusters based on new spectroscopic 
redshifts (five clusters) or additional \chandra\ observations (five clusters). 
We refer the reader to A11 and B11 for additional details on these 
X-ray observations and the analysis of the associated data; the X-ray data here are identical to that used by B11.

From the X-ray data on this 14-cluster sample, density and temperature profiles were derived for use in our cosmological analysis in Section \ref{sec:cosmo}.  
This was done by calculating \Txtheta\ and \Mgr\ (allowing the calculation of $\Yx(r)$ given a reference cosmology) from the X-ray observations of each cluster.  
Here $r$ corresponds to a physical radius in the cluster,  \Mgr\ is the gas mass, \Tx\ is the core-excised X-ray temperature, and \Yx\ is the product of \Mg\ and \Tx. 

\subsubsection{\rosat{} Counterparts}

A number of cluster candidates are found to be associated with sources in the
\rosat\ Bright or Faint Source Catalog \citep{voges99,voges00}.
For each of these, Table \ref{tab:rosat} lists intrinsic X-ray fluxes
and rest-frame luminosities in the 0.5--2.0~keV band, inferred from
the \rosat{} count rates. 
The luminosities assume a reference cosmology chosen to match A11, 
who assumed a {\sl WMAP}7+BAO+{\sl $H_0$} $\Lambda$CDM preferred cosmology
 with $\Omega_M = 0.272$, $\Omega_\Lambda = 0.728$ and $H_0 = 70.2~$km$~$s$^{-1}~$Mpc$^{-1}$ \citep{komatsu11}. 
 The absorbing column density of Galactic hydrogen towards each cluster was accounted for using
the H$_\mathrm{I}$ survey of \citet{kalberla05}, and the necessary
redshift- and temperature-dependent K-corrections were performed using ICM temperature estimates based on the
SPT signal to noise for each cluster for a simple power-law fit to
the A11 data.\footnote{We note, however, that the resulting flux
and luminosity estimates are largely insensitive to the temperatures used.
For example, adopting the temperature--luminosity relation of
\citet{mantz10b} results in luminosities that differ by $\sim 2\pm2\%$, far less
than the typical statistical uncertainty in the \rosat{} count rates.}

These \rosat-derived observables are reported only to provide further confirmation of these clusters;  
we emphasize that these results are \emph{not} used in the cosmological analysis.
Rather, only the X-ray observables from the 14-cluster \chandra \ and \xmm \ dataset from A11 and B11 are used in 
the cosmological analysis.

\begin{deluxetable*}{lcc  c | c c c}
\centering
\tablecaption{\label{tab:rosat}\rosat\ counterparts} \small
\tablehead{
SPT ID & \rosat\ ID & Offset & $z$ & \rosat{} counts & $F_X $ & $L_X $  \\
 & & ($^{\prime\prime}$)  & &(${\rm s}^{-1}$) & $(10^{-13}\, {\rm ergs\,cm}^{-2} {\rm s}^{-1})$ & $(10^{44}\, {\rm erg\,s}^{-1})$ \\
}

\startdata
\input{tab_xray.tex}
\enddata

\tablecomments{ Cluster candidates coincident with sources in the \rosat\ bright or faint source catalogs \citep{voges99,voges00}. 
We define a match if a candidate is within 5$^\prime$
of a cluster candidate at $z \le 0.3$ or within 2$^\prime$ of a candidate at $z > 0.3$.  
For each source, we estimate the X-ray luminosity and flux based on the measured redshift, position on the sky, and \rosat\ X-ray photon counts. 
Clusters marked with a `*' also have $Y_X$ estimates from XMM or Chandra presented by A11 and B11. 
Note that SPT-CL J0311-6354 is coincident with 1ES0310-64.0, but not a \rosat{} source. 
We also quote the cluster redshift used in this work (see \S\ref{sec:optical}). 
We include error bars for red sequence redshifts, but not spectroscopic redshifts.
}
\end{deluxetable*}

\section{Cluster Catalog}
\label{sec:catalog}

In Table \ref{tab:catalog}, we present the complete list of galaxy cluster candidates from  $720$ deg$^2$  of sky surveyed by the SPT. 
The catalog includes \ncand\ galaxy cluster candidates with detection significance, $\xi \ge 4.5$. 
Using optical/NIR follow-up data (see \S\ref{sec:optical}), we have determined redshifts 
for \nconfirmfourf\ of the SPT-selected galaxy cluster candidates.
The median redshift of the sample is $z=\medianz$. 
The left panel of Figure \ref{fig:zhist} shows the redshift histogram of our cluster sample. 
The right panel shows SZ detection significance versus redshift for each cluster with an estimated 
redshift.

We search for galaxy clusters published in other catalogs within 2~arcmin of every candidate
reported in Table \ref{tab:catalog} and within 5~arcmin of any candidate in Table \ref{tab:catalog} 
at $z \le 0.3$.  We query the SIMBAD\footnote{http://simbad.u-strasbg.fr/simbad} and 
NED\footnote{http://nedwww.ipac.caltech.edu/} databases, and we 
manually search more recently published cluster catalogs such as the PLCKESZ 
\citep{planck11-5.1a} and ACT-CL \citep{marriage11b} catalogs.  All matches within the 
appropriate radius are listed in Table \ref{tab:otherid}; whether the associations are physical 
or random superpositions is discussed in S12.

The optically confirmed, SZ-selected galaxy clusters are found to be massive, with a sharp mass cutoff at approximately $M_{500}=\mthreshold$\ at $z = 0.6$. 
We define $M_{500}$ as the mass within a sphere of radius $r_{500}$, defined as the radius at which the density is 500 times the critical density. 
The exact mass cutoff depends on the field and cluster redshift. 
We discuss mass estimates for the clusters in \S\ref{sec:mass}, and we show the estimated masses versus redshift in the left panel of Figure \ref{fig:mass}. 
The most massive cluster is SPT-CL J2106-5844 at $z = 1.1320$ with a mass of $M_{500}=8.39 \pm 1.68 \times10^{14}\,M_\odot\,h_{70}^{-1}$. 
This is  the most massive cluster at $z > 1$ currently known. 
\citet{foley11} showed that although this cluster is rare, it is not in significant tension with the \lcdm{} model.
The least massive is SPT-CL J2007-4906 at $z = 1.25\pm 0.11$ with $M_{500}=2.11 \pm 0.82 \times10^{14}\,M_\odot\,h_{70}^{-1}$. 
The median mass of the sample is \medianm. 

We compare the mass and redshift distribution of this SPT cluster catalog to cluster catalogs from the \rosat{} and  \planck{} all-sky surveys in Figure~\ref{fig:masscomp}. 
For the \rosat{} all-sky survey, we show 917 clusters taken
from the NORAS, REFLEX, and MACS cluster catalogs, as given in the MCXC compilation \citep{piffaretti11}.
 We use the redshift and mass estimates reported by \citet{piffaretti11}, where the masses were estimated from the X-ray luminosity-mass relation. 
We also show the 155 out of 189 galaxy clusters in the  Planck-ESZ cluster catalog \citep{planck11-5.1a} that have counterparts in the MCXC compilation.
The plotted masses and redshifts for these clusters are taken from the MCXC compilation. 
The mass estimates for the SPT clusters are described in \S\ref{sec:mass}. 
The selection function of the SPT catalog is nearly independent of redshift. 
In fact, the minimum 
mass drops slightly with redshift as the angular size of galaxy clusters 
decreases, becoming better matched to the SPT beam and less confused by 
primary CMB fluctuations. 
This reduction in size with increasing redshift has the opposite effect on the \planck{} SZ survey 
due to the \planck{} satellite's larger beam size ($7^\prime$ at 143\,GHz). 
Beam dilution reduces the \planck{} satellite's signal-to-noise on high 
redshift clusters, while the outstanding frequency coverage makes it 
possible to subtract the primary CMB on large angular scales and recover 
the SZ signal from low-redshift galaxy clusters.
Finally, the \rosat{} cluster mass threshold rises with redshift due to cosmological dimming of the X-ray flux, crossing over the SPT selection function around $z\sim 0.3$.

The catalog presented here is expected to be 95\% pure for detection significance $\xi \ge 5$ and 71\% pure for detection significance $\xi \ge 4.5$. 
This agrees well with the actual optical and NIR confirmation rate. 
From \S\ref{subsec:simpurity}, we expect 59 (6.4) candidates to be false above a detection significance of 4.5 (5). 
We find \nfalsefourf{} (\nfalsefive{}) candidates do not have optical counterparts, which is in excellent agreement with the expected number of false detections.

\begin{deluxetable*}{ llccc}
\centering
\tablecaption{\label{tab:otherid}Clusters with matches in other catalogs} \small
\tablehead{
SPT ID & First ID, ref. & All catalogs with match & $z$ & Lit.~$z$, ref.  \\
}
\startdata
\input{tab_other.tex}
\enddata
 \tablenotetext{A}{SPT-CL catalog.  W11}
 \tablenotetext{B}{PLCKESZ catalog.  \citet{planck11-5.1a}}
 \tablenotetext{C}{ACO catalog.  \citet{abell89}}
 \tablenotetext{D}{APMCC catalog.  \citet{dalton97}}
 \tablenotetext{E}{[DBG99] catalog.  \citet{degrandi99}}
 \tablenotetext{F}{[DEM94] catalog.  \citet{dalton94}}
 \tablenotetext{G}{REFLEX catalog.  \citet{bohringer04}}
 \tablenotetext{H}{\citet{struble99}}
 \tablenotetext{I}{\citet{ebeling96}}
 \tablenotetext{J}{Sersic catalog.  \citet{sersic74}}
 \tablenotetext{K}{Stromlo catalog.  \citet{duus77}}
 \tablenotetext{L}{SPT-CL catalog.  \citet{staniszewski09}}
 \tablenotetext{M}{SPT-CL catalog.  V10}
 \tablenotetext{N}{ACT-CL catalog.  \citet{marriage11b}}
 \tablenotetext{O}{H10}
 \tablenotetext{P}{SCSO catalog.  \citet{menanteau10}}
 \tablenotetext{Q}{\citet{brodwin10}}
 \tablenotetext{R}{[QW] catalog.  \citet{quintana90}}
 \tablenotetext{S}{ClG catalog.  \citet{fetisova81}}
 \tablenotetext{T}{\citet{suhada10}}
 \tablenotetext{U}{\citet{buckleygeer11}}
\tablecomments{ Cluster candidates coincident with galaxy clusters identified in other catalogs. 
We define a match if a candidate is within 5$^\prime$ (2$^\prime$) of an identified cluster for clusters at $z < 0.3$ ($z > 0.3$ or unconfirmed). 
For each match, we report the name under which the cluster was first reported and all catalogs which include the cluster.  See S12 for a discussion of physical association vs.~random superposition for these matches.
We also quote the cluster redshift used in this work---either the photometric redshift estimated in S12 or a spectroscopic redshift obtained from followup observations  or the literature. 
 We include error bars for red sequence redshifts but not spectroscopic redshifts.  
In the last column, we quote   a redshift from the literature if available. 
 Error bars are not reported for literature redshifts; two (four) significant digits are used if the literature redshift is photometric (spectroscopic). 
}
\end{deluxetable*}

\subsection{Cluster Candidates in the Point-source-masked Regions}
\label{sec:psource}

As discussed in Section \ref{sec:extract}, any cluster detections within $8'$ of 
an emissive point source detected above $5 \sigma$ at 150~GHz are rejected. 
We do this because residual source flux or artifacts due
to the masking of these point sources can cause spurious decrements when the maps are filtered.
A total area of $\sim$50 out of 770 square degrees ($\sim$6.5$\%$) was excluded from cluster finding for this reason. 
This conservative procedure is appropriate for constructing a cluster catalog with a clean, easy-to-define
selection function and a mass-observable relation with minimal 
outliers.  
However, it is  likely that several massive clusters will lie within the exclusion 
region, and some of those clusters might be only minimally affected by the nearby
emissive source.  If we assume no spatial correlation between sources and clusters,
we would expect roughly eight missed clusters above $\xi = 5$.

As in W11, we re-ran the cluster-finding algorithm on all the fields
used in this work with only the very brightest 
sources masked.  For this work we used a bright-source threshold of 
$S_{150 \mathrm{GHz}}>100$~mJy, compared to the normal threshold
of $\sim$6~mJy, resulting in a total masked area of $< 3$ square degrees.  
Each detection with $\xi \ge 5$ from the originally masked area was visually 
inspected (below the $\xi=5$ threshold, it becomes too difficult to distinguish visually between 
real clusters and artifacts), and the vast majority were rejected as obvious point-source-related artifacts.  
Some detections, however, did appear to be significant SZ decrements only
minimally affected by the nearby source.  These objects are listed in Table
\ref{tab:missed}.
We find six objects above $\xi = 5$, consistent within Poisson uncertainties
with the expected number.  One of these objects, SPT-CL~J2142-6419, 
was also identified in the auxiliary detection procedure in W11.  Two of the 
six objects (SPT-CL~J2154-5952 and SPT-CL~J2154-5936)
are unusually close to one another on the sky ($16\farcm 5$ 
separation), but visual inspection shows nothing out of the ordinary about either
candidate beyond its proximity to an emissive source.

We have not yet attempted to obtain redshifts for these six cluster candidates, and 
they are not included in the cosmological analysis or in the total number of 
candidates quoted in the rest of the text.  We perform the same search for 
counterparts to these six candidates in other galaxy cluster catalogs as 
we do for the main sample.  We find no galaxy cluster matches, 
though we do find X-ray sources within 5~arcmin of
SPT-CL~J0334-6008 and within 2~arcmin of SPT-CL~J0434-5727, 
SPT-CL~J0442-5905, and SPT-CL~J2154-5936.

\begin{deluxetable}{ lcccc}
\centering
\tablecaption{\label{tab:missed}Cluster candidates above $\xi=5$ in the source-masked area} \small
\tablehead{
SPT ID & RA & DEC & \nSN & $\theta_c$  \\
}
\startdata
\input{tab_missed.tex}
\enddata
\tablecomments{ Cluster candidates identified in a non-standard cluster-finding
analysis, in which only the very brightest ($> 100$~mJy) point sources are masked
(see text for details).
Only candidates from the area masked in the standard analysis are listed here.  
These candidates are not included in the cosmological analysis or in the total number
of candidates quoted in the text.
}
\end{deluxetable}

\section{Cosmological modeling}
\label{sec:cosmo}

In this section, we briefly review the method presented by B11 to use the combination of an SZ-selected cluster catalog and X-ray follow-up observations to investigate cosmological constraints; we refer the reader to B11 for a complete description. 
We also present a slightly modified algorithm  to treat fields of varying depths.

We use Monte Carlo Markov chain (MCMC) methods to determine parameter constraints. 
As outlined by B11, we have extended CosmoMC\footnote{http://cosmologist.info/cosmomc/} 
\citep{lewis02b} to simultaneously fit the SZ and X-ray cluster observable-mass relations 
while also varying cosmological parameters. 
We include all \nSN{} $> 5$ and $z > 0.3$ cluster candidates in this catalog (\ncosmo\ clusters) in the cosmological analysis as well as the X-ray observations of 14 SPT-selected clusters described by A11 and B11. 
These data will be referred to as \sptcl{}.

In addition to the cluster data, 
some MCMCs include CMB data from WMAP7 and SPT \citep{komatsu11,keisler11}. 
In some cases, we also add measurements of the BAO feature using SDSS and 2dFGRS data \citep{percival10}, low-redshift measurements of $H_0$ from the Hubble Space Telescope \citep{riess11}, or measurements of the luminosity-distance relationship from the Union2 compilation of 557 SNe \citep{amanullah10}.
Finally, we sometimes use a BBN prior on the baryon density from measurements of the abundances of deuterium \citep{kirkman03}. 
In all cases, we set the helium abundance based on the predictions of BBN \citep{hamann08}. 

\subsection{X-ray scaling relations}

Following \citet{vikhlinin09} and B11, we use \Yx\ as an X-ray proxy for cluster mass, \mass.  We assume a \yxm\ relation of the form
\begin{equation}
\frac{\mass}{ 10^{14} M_{\odot} / h } = \left(A_X h^{3/2}\right) \left( \frac{Y_X}{3 \times 10^{14} M_{\odot}\, {\rm keV}} \right)^{B_{X}} \left(\frac{H(z)}{ H_0}\right)^{C_{X}},
\label{eqn:yxm}
\end{equation}
parameterized by the normalization $A_{X}$, the slope $B_{X}$, the redshift evolution $C_{X}$, and a log-normal scatter $D_{X}$ on $Y_X$. 
We express the mass in units of  $M_{\odot}/h$ to match the \snm\ relation in Section \ref{sec:szscale}. 
 For our cosmological analysis, we 
assume the same Gaussian priors on the scaling relation parameters as B11.  
The priors are motivated by constraints from X-ray measurements by \citet{vikhlinin09b} and simulations. 
The Gaussian priors are $A_{X} = 5.77 \pm 0.56$,  $B_X=0.57\pm0.03$, $C_X=-0.4\pm0.2$, and $D_X=0.12\pm0.08$. 
For the cosmological results in this paper, only the uncertainty on $A_{X}$ matters; we have tested fixing the other parameters to their central values and find essentially identical results.

\subsection{SZ scaling relations}
\label{sec:szscale}

As in \citetalias{vanderlinde10} and B11, we estimate galaxy cluster masses according to an SZ signal-to-noise to mass scaling relation. 
Following those works, we introduce the 
unbiased significance, $\zeta$, since the relation between \nSN\ and 
halo mass is complicated by the comparable effects  of intrinsic scatter and instrumental noise. 
The unbiased significance is defined to be the average detection signal-to-noise of a simulated cluster, measured across many 
noise realizations, and related to the detection significance \nSN\ as follows:
\begin{equation}
 \zeta = \sqrt{\langle\nSN\rangle^2-3}
\label{eqn:nsn}
\end{equation}
at $\nSN>2$. 
The detection significance \nSN\ is maximized across possible cluster positions and filters scales, effectively  
 adding three degrees of freedom to the fit. 
 The unbiased significance $\zeta$  
 removes this maximization bias.

The specific form of the scaling relation is: 
\begin{equation}
\label{eq:mass_scaling}
\zeta = A_{SZ} \left( \frac{\mass}{3 \times 10^{14} M_{\odot} h^{-1}} \right)^{B_{SZ}} \left(\frac{H(z)}{H(0.6)}\right)^{C_{SZ}},
\end{equation}
where $A_{SZ}$ is a normalization, $B_{SZ}$ a mass evolution, and $C_{SZ}$ a redshift 
evolution. 
The method to go from simulations to an SZ signal-to-noise to mass scaling relation is described in more detail by \citetalias{vanderlinde10}. 

As described more fully in \citetalias{vanderlinde10}, this scaling relation is based on SZ simulations of approximately $4000\,$deg$^2$ of sky. 
The simulations used in this work are described in \S\ref{sec:sims}. 
The intrinsic scatter, $D_{SZ}$, was measured to be 24\%. 
The main uncertainty for cosmological purposes is on the mass normalization, $A_{SZ}$, which is assumed to be uncertain at the 30\% level.

Unlike \citetalias{vanderlinde10}, this analysis includes fields with substantially different noise levels. 
We have repeated the simulations (see \S\ref{sec:sims}) on each field, and find the main effect is an overall rescaling of the expected SZ signal-to-noise for a given cluster mass, i.e., a change to $A_{SZ}$. 
There is a slight change to the redshift evolution between fields as well, but neglecting this results in an additional percent level scatter which is completely negligible given the overall 24\% scatter in the scaling relation. 
We have also checked the simulations by adding known cluster profiles to the real maps, applying the cluster-finding algorithm and checking the recovered signal-to-noise. 
This semi-analytic test agrees well with the results of the simulations. 
We apply a fixed rescaling of $A_{SZ}$ to each field, as tabulated in Table \ref{tab:rescale}. 
The normalization of the rescaling is chosen such that the \five{} field is unity. 
We use simulations of all five fields to estimate the parameters $A_{SZ}$, $B_{SZ}$,  $C_{SZ}$, and $D_{SZ}$ for the combined scaling relation, and determine values of  6.24, 1.33, 0.83, and 0.24 respectively. 
Uncertainties in the SZ modeling lead to significant systematic uncertainties on these scaling relation parameters. 
Following \citetalias{vanderlinde10}, we apply conservative 30\%, 20\%, 50\%, and 20\% Gaussian uncertainties to $A_{SZ}$, $B_{SZ}$,  $C_{SZ}$, and $D_{SZ}$, respectively.

\begin{deluxetable}{ l c c}
\centering
\tablecaption{\label{tab:rescale}SZ-Mass normalization per field} \small
\tablehead{
Name & Year & Scaling factor\\
}
\startdata
\five & 2008 & 1.00\\
\twthree &  2008 & 1.01\\
\three  & 2009 &  1.25\\
\twonefif  & 2009 &  1.09\\
\twonesix  & 2009 &  1.31\\
\enddata

\tablecomments{ The estimated scaling factors for the mass normalization $A_{SZ}$ for each field. These factors correct for the different noise levels in each field. }
\end{deluxetable}

\subsection{Cluster likelihood function}

We have written a module extension to CosmoMC to calculate the cluster likelihood function. 
In essence, this module uses the Cash statistic \citep{cash79} to compare the observed number counts to a known Poisson distribution at each step in the MCMC. 
The method closely mirrors that presented by  B11, to whom we refer the reader for a complete description. 
Briefly, we use the Tinker mass function \citep{tinker08} to calculate the mass function based on the cosmological parameters and associated matter power spectra estimated by CAMB \citep{lewis00} at 20 logarithmically spaced redshifts between $0 < z < 2.5$. 
The mass function is calculated for an over-density of 500 times the critical density. 
Using the scaling relation parameters at that step of the MCMC, the mass function is translated from the native $M_{500}$-$z$ space into the three-dimensional observable space, with axes corresponding to the SZ detection significance $\nSN$, the X-ray parameter \Yx, and the optically derived redshift $z$. 
The observed number counts are compared to the expectation values in this three dimensional space to evaluate the likelihood for that step of the MCMC.

There are two differences between the likelihood function used in this work and that presented by B11. 
The most significant of these is the field-dependent SZ scaling relation described in \S\ref{sec:szscale}. 
In practice, this means that the above calculation is done separately for each of the five fields, and the resulting log likelihoods are summed. 

The treatment of unconfirmed cluster candidates is the second, more minor difference between this work and B11. 
B11 left unconfirmed clusters out of the analysis; this is appropriate given the extremely high redshift lower limit on the single unconfirmed (and almost certainly false) cluster candidate in that cluster sample. 
A more rigorous treatment  includes the likelihood of each unconfirmed candidate,
using the expectation value of the candidate being either a higher redshift cluster or a false detection. 
This expectation value is the sum of the expected number of false detections at a given detection significance and a redshift-dependent selection function convolved by the mass function. 
In practice, the treatment of unconfirmed clusters is nearly negligible since the $\SN > 5$ sample used to derive cosmological constraints has high purity and the precision of the cosmological constraints is currently limited by the systematic mass calibration uncertainty.

With this in mind, we make two simplifying approximations in our implementation.
First, we neglect any cosmological dependence in the false detection rate -- the simulations used to calculate the false detection rate are only run for one cosmological model. 
This effect should be negligible since the CMB and foreground power levels are well-known. 
Second, we treat the redshift selection function as a Heaviside function at the quoted redshift limit for that cluster. 
The chance of detecting a cluster out to this redshift is nearly unity with the current optical and NIR observations. 
We have tested shifting the Heaviside function to $z > 1.5$ or $z > 2$ and find no impact on the cosmological constraints. 
This can be understood intuitively because (a) the overall purity is high, and (b) the expected number of unconfirmed, real, high-redshift clusters is small compared to the expected number of false detections.

\section{Cosmological implications}
\label{sec:constraints}

We present cosmological constraints from the SPT cluster sample in this section. 
 The main results  are tabulated in Table \ref{tab:cosmo}. 
We first consider the baseline, six-parameter \lcdm{} model, and highlight the improvement in parameter constraints for the current catalog compared to the smaller B11 catalog. 
The  uncertainty in the cosmological analysis is dominated by the systematic cluster mass uncertainty; the mass calibration is based on the same X-ray data and  $Y_X - M$ scaling relation  used in the analysis of B11. 
The $Y_X - M$ scaling relation was observationally measured
using X-ray measurements of the total mass derived assuming
hydrostatic equilibrium, with the absolute calibration cross-checked
by weak-lensing-based mass estimates.
We also compare the observed and expected cluster abundances, and we estimate the masses of each galaxy cluster.
We next examine cosmological constraints for two extensions beyond a \lcdm{} model,  testing the ability of the cluster sample to constrain the dark energy equation of state and the sum of the neutrino masses. 
These two model extensions are degenerate in the current cluster data; we choose to look at independent constraints on each while fixing the other parameter to the  \lcdm{} baseline value. 
Finally, we discuss prospects for improving the mass calibration and thereby realizing the full potential of SZ-selected galaxy clusters as cosmological probes.

\subsection{\lcdm\ constraints}

\begin{figure}[]
\centering
\includegraphics[width=.45\textwidth]{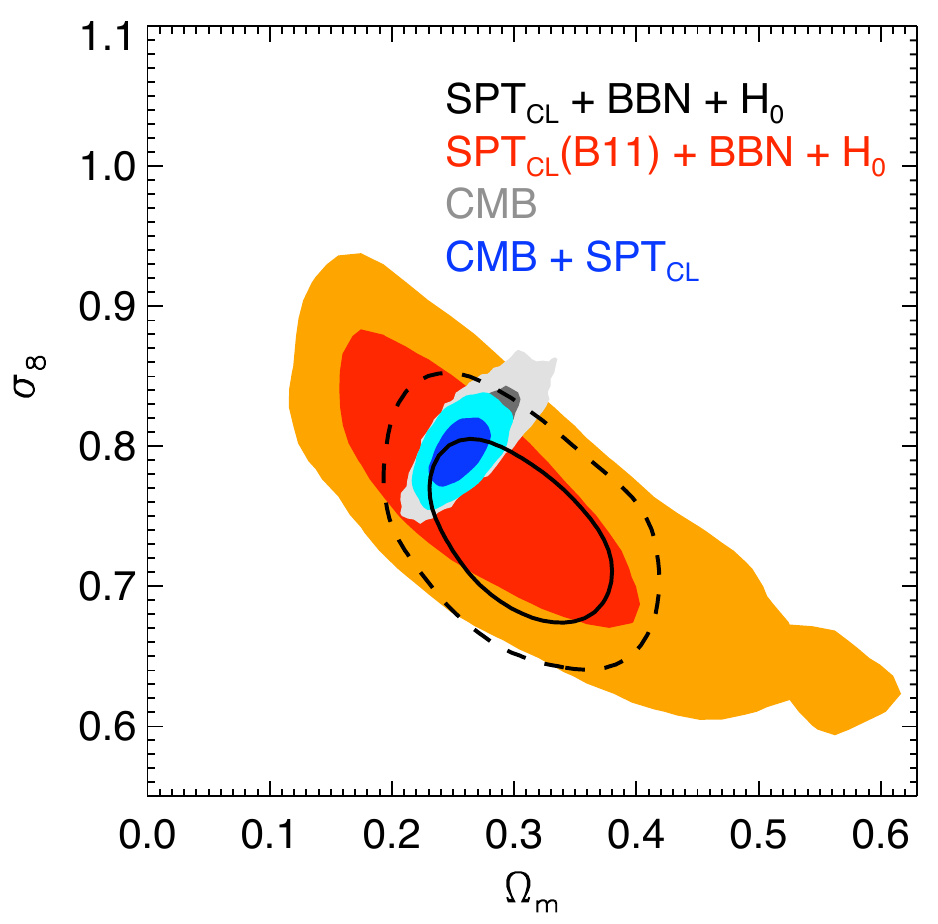}
  \caption[]{68\% and 95\% likelihood contours in the $\Omega_m - \sigma_8$ plane for different datasets. 
  The additional clusters in this catalog  reduce  the allowed parameter volume by a factor of two compared to B11 when considering the \sptcl+BBN+H$_0$ data. 
  However,   the linear combination showing the most improvement is already well constrained by the CMB data. 
  Therefore as predicted by B11, the additional clusters will not substantially improve constraints for the CMB+\sptcl{} data until the mass calibration is improved.
      \\ }
\label{fig:s8om_lcdm}
\end{figure}

In a \lcdm\ model, cluster samples primarily constrain  $\sigma_8$ and $\Omega_m$ (see e.g., B11, \citealt{rozo10}). 
As was done by B11, we look at ``cluster-only" constraints based on \sptcl+BBN+H$_0$ with the reionization optical depth fixed to $\tau = 0.08$. 
The external data and $\tau$ prior are required since cluster abundances are insensitive to several \lcdm{} parameters, including $\tau$. 
We see a substantial improvement in the \sptcl+BBN+H$_0$ constraints with the expanded cluster catalog from this work; the allowed likelihood volume is reduced by approximately a factor of two (compare the filled red/orange contours and black contours in Fig. \ref{fig:s8om_lcdm}).

Adding the new SPT cluster sample to the WMAP7 and SPT CMB power spectrum data improves the constraints on $\Omega_m, \Omega_c h^2, \sigma_8, {\rm and\,} h$ by roughly a factor of 1.5 over the CMB alone.  
The constraints are listed in the first two columns of Table~\ref{tab:cosmo}. 
The cluster data modestly tightens constraints on the amplitude of the primordial power spectrum as well. 
The uncertainty on the amplitude is reduced by 24\% from $ln (10^{10}A_s) = 3.196 \pm 0.042$ to $3.176 \pm 0.034$. 

However, these constraints are only marginally better than those presented by B11. 
As shown in Fig.~\ref{fig:s8om_lcdm},  the \sptcl{} constraints in the $\Omega_m-\sigma_8$ plane are most improved along a direction  well-constrained by the CMB data. 
For constraints in the perpendicular direction, the \sptcl{} data are limited by the current mass calibration uncertainty, determined from the $Y_X - M$ scaling relation which is unchanged from B11. 
  A better mass calibration will be essential to realize the full potential of cosmological 
constraints from galaxy clusters.

\subsubsection{Comparison to expected number counts}

We can compare the observed number of galaxy clusters with the number expected 
for a given cosmological model. 
We compare the high-purity sample of  \ncosmo{} cluster candidates at $z > 0.3$ and $\nSN > 5$ to the expected number counts for two cases with a \lcdm\ cosmology. 
The first case uses only non-cluster data (CMB+BAO+SN+H$_0$) with the scaling relations allowed to vary across the conservatively wide simulation-based prior. 
The symmetric $1\,\sigma$ range around the median is [121, 805] candidates and the $2\,\sigma$ range is [37, 2004] candidates. 
The likelihood peaks near 130 candidates. 
In the second case, we add the cluster catalog data while allowing the scaling relations to vary. 
As would be expected, this MCMC leads to a tighter predicted range of candidate counts with roughly Poisson scatter around the observed number counts. 
The $1\,\sigma$ range is [92, 111] candidates with the median at 101 objects. 
The difference between the two cases is primarily due to the range of scaling relation parameters explored. 
We do not see significant tension with the observed cluster counts in either case. 

\subsubsection{Mass estimates}
\label{sec:mass}

We present mass estimates based on the posterior probability distributions for all optically confirmed clusters  in Table \ref{tab:catalog}. 
In all cases, we quote $M_{500}$, as defined in \S\ref{sec:catalog}.
For the 15 clusters with X-ray data from A11, these are joint X-ray and SZ  mass estimates. 
Only the SZ data are used for the other clusters. 
We calculate a probability density function on a mass grid at each point in the CMB+BAO+SN+H$_0$+\sptcl\ parameter chain for a \lcdm\ cosmology. 
The allowed  \lcdm\ parameter ranges for this data set are essentially unchanged from the CMB+\sptcl{} data set.
These probability density functions are combined to obtain a mass estimate that has been fully marginalized over all cosmological and scaling relation parameters.

\begin{deluxetable*}{ c | c c | c c | c c  }
\centering
\tablecaption{\label{tab:cosmo}Cosmological constraints} \small
\tablehead{
& \multicolumn{2}{c}{\lcdm}& \multicolumn{2}{c}{wCDM}& \multicolumn{2}{c}{$\sum m_\nu$}\\
\\
& CMB & +\sptcl\ & \wset\ & + \sptcl\ & \nuset\ & + \sptcl\ \\
}
\startdata
$\Omega_c h^2$ & $  0.1109 \pm   0.0048$ & $  0.1086 \pm   0.0031$ &$  0.1140 \pm   0.0041$ &$  0.1104 \pm   0.0029$ &$  0.1113 \pm   0.0030$ &$  0.1113 \pm   0.0025$   \\  
$\sigma_8$ & $   0.808 \pm    0.024$ & $   0.798 \pm    0.017$ &$   0.840 \pm    0.038$ &$   0.807 \pm    0.027$ &$   0.775 \pm    0.041$ &$   0.766 \pm    0.028$   \\  
$\Omega_m$ & $   0.267 \pm    0.026$ & $   0.255 \pm    0.016$ &$   0.269 \pm    0.014$ &$   0.262 \pm    0.013$ &$   0.274 \pm    0.016$ &$   0.275 \pm    0.015$   \\  
$H_0$ & $   70.71 \pm     2.17$ & $   71.62 \pm     1.53$ &$   71.20 \pm     1.49$ &$   71.15 \pm     1.51$ &$   69.83 \pm     1.36$ &$   69.76 \pm     1.31$   \\  
\\
$w$ & & & $  -1.054 \pm    0.073$ & $  -1.010 \pm    0.058$ & & \\
\\
$\sum m_\nu$ (95\% CL) & & & & & $< 0.44$ & $< 0.38$ \\ 

\enddata
\tablecomments{ Cosmological constraints for three models with and without the SPT cluster sample.}
\end{deluxetable*}

\subsection{Dark energy equation of state}

We next examine cosmological constraints in a  \wCDM\ cosmology.
This model introduces the dark energy equation of state, $w$,   as a free parameter 
(in the \lcdm \ model, $w$ is fixed to $-1$). 
The equation of state remains constant with time. 
The cluster abundance and the shape of the mass function depend on $w$ 
through its effect on the expansion history of the Universe and the growth of structure.

\begin{figure*}[t]\centering
\includegraphics[width=\textwidth]{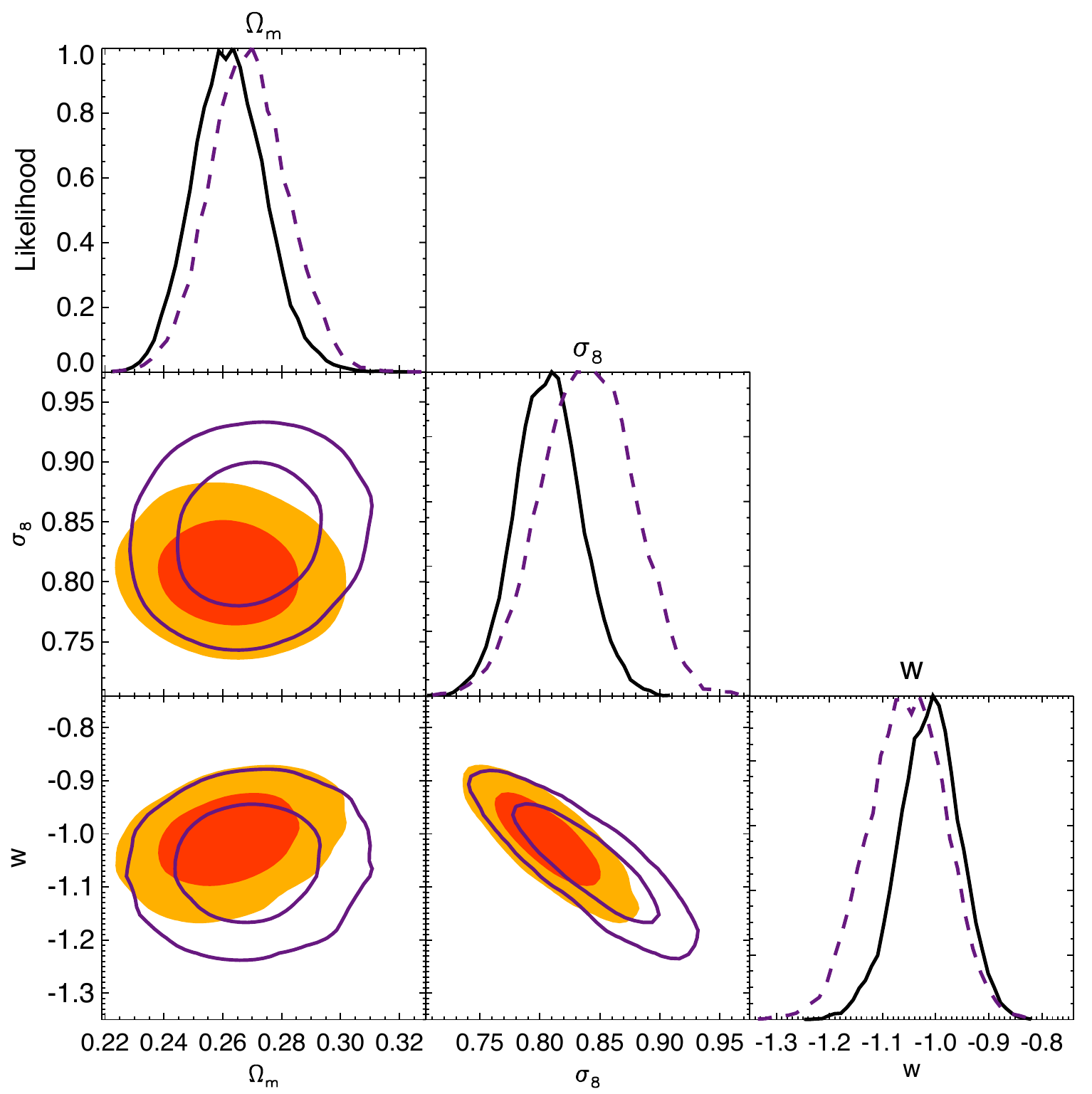}
  \caption[]{Assuming a \wCDM\ cosmology, the constraints on $\Omega_{m}$, $\sigma_8$, 
  and $w$.  The plots along the diagonal are the one-dimensional 
  marginalized likelihood.  The off-diagonal plots are the two-dimensional marginalized constraints showing the 
  68\% and 95\% confidence regions.  We show the constraints for the  
  \wset{} (purple line contours and  dashed lines), and \wset{}+\sptcl{} (filled contours and black, solid lines) data sets.
  Including the \sptcl\ data improves the constraints on $\Omega_{m}$, $\sigma_8$, and $w$ 
  by factors of 1.1, 1.4, and 1.3 respectively.
        \\
}
\label{fig:wcdm}
\end{figure*}

The best external constraints on $w$ come from a combination of the CMB, BAO, H$_0$, and SNe data. 
Adding the SPT cluster sample to this data set reduces the uncertainty on the dark energy equation of state by a factor of 1.3 to give \wWCDM. 
This value is completely consistent with a cosmological constant and within $1\,\sigma$ of the no-cluster constraint of $w=-1.054 \pm 0.073$. 
These results are shown in Figure \ref{fig:wcdm} and tabulated in Table \ref{tab:cosmo}.

The cluster data also aid in the measurement of the dark matter density and $\sigma_8$.
The addition of clusters moves the preferred cold dark matter density down by nearly $1\,\sigma$ from $\Omega_c h^2 = 0.1140 \pm 0.0041$ to $\Omega_c h^2 = 0.1104 \pm 0.0029$. 
As would be expected, the amplitude of the matter power spectrum also drops from $\sigma_8 = 0.840 \pm 0.038$ to \sigmaeightWCDM. 
The uncertainties on both parameters are reduced by a factor of 1.4 with the addition of the SPT cluster data. 

We can compare the wCDM results to those reported by B11 based on fewer clusters but the same X-ray data and mass calibration uncertainty. 
B11 report $\sigma_8 = 0.793 \pm 0.028$ and $w = -0.973 \pm 0.063$ for the   CMB + BAO + SNe + \sptcl{}(B11) data. 
In this analysis, the median $\sigma_8$ and $w$ values shift by $\sim$$\,0.5\,\sigma$ relative those presented by B11, and the 
uncertainties tighten slightly.
These changes are primarily due to including the local measurement of H$_0$ in the constraints, rather than the additional clusters. 
We also ran chains without H$_0$ to parallel the B11 treatment and both differences effectively disappear.

\subsection{Massive neutrinos}

\begin{figure*}[t]\centering
\includegraphics[width=\textwidth]{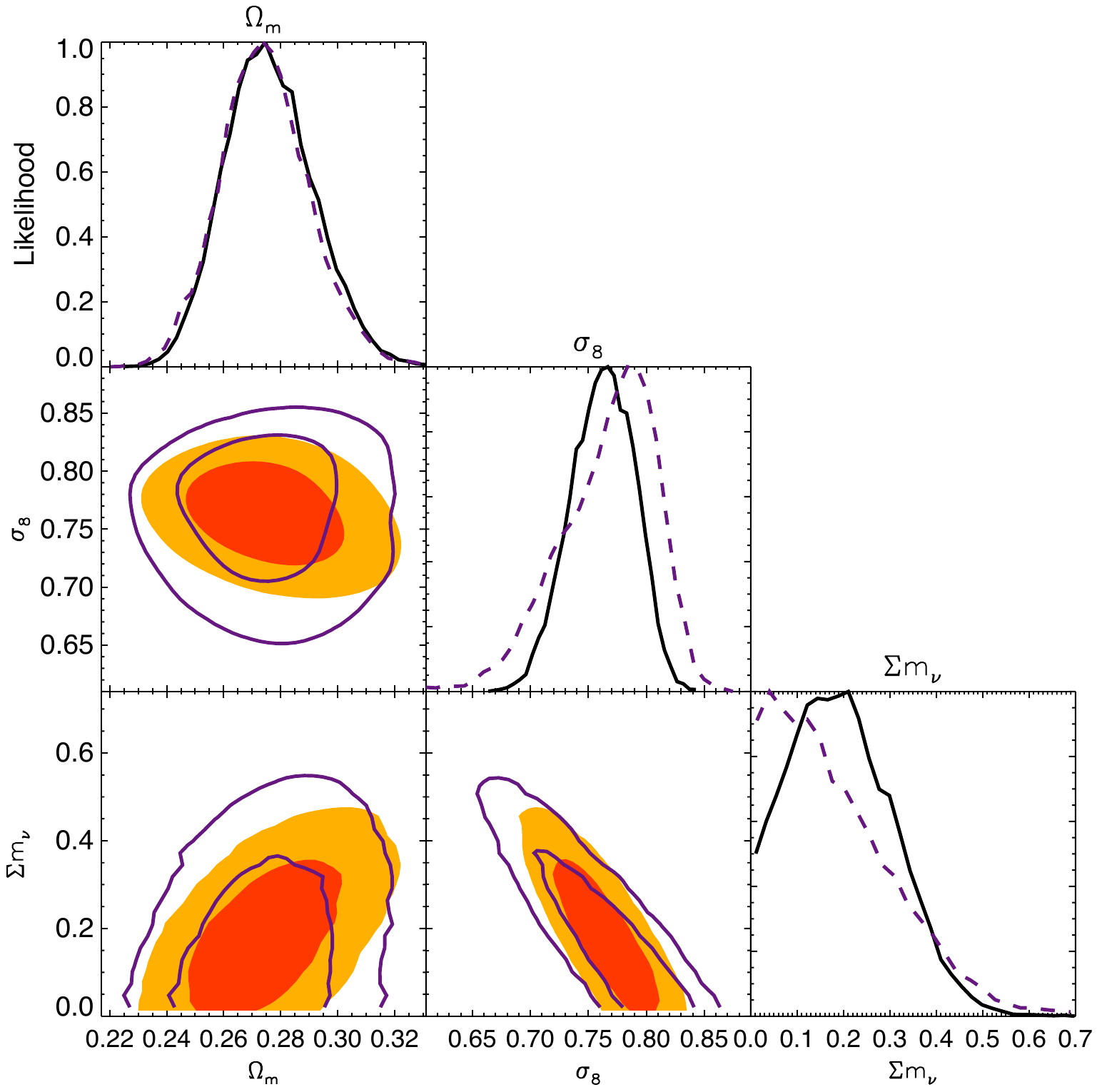}
  \caption[]{Assuming a \lcdm{} + massive neutrino cosmology, the constraints on $\Omega_{m}$, $\sigma_8$, 
  and $\sum m_\nu$.  The plots along the diagonal are the one-dimensional 
  marginalized likelihood.  The off-diagonal plots are the two-dimensional marginalized constraints showing the 
  68\% and 95\% confidence regions.  We show the constraints for the  
  \nuset{} (purple line contours and  dashed lines), and \nuset{}+\sptcl{} (filled contours and black, solid lines) data sets.
  The \sptcl\ data leads to a small preference for positive neutrino masses  with $\sum m_\nu = 0.17 \pm 0.13$\,eV; 
  the  95\% CL upper limit on the neutrino masses is reduced from  $\sum m_\nu < \mnunoclus$\,eV to $\sum m_\nu < \mnu$\,eV. 
        \\
}
\label{fig:mnu}
\end{figure*}

The second extension to a \lcdm\ model that we consider is a non-zero sum of neutrino masses, $\sum m_\nu \ge 0$. 
Non-zero neutrino masses are well-motivated by the measured mass differences in neutrino oscillation experiments (e.g., \citealt{ahmad02,eguchi03,ashie04}). 
For CMB + H$_0$ + BAO, neutrino masses are highly degenerate with $\sigma_8$, as shown in Figure \ref{fig:mnu}. 
Cluster abundances are an independent measure of local structure ($\sigma_8$), and thereby enable better constraints on the sum of the neutrino masses. 
In this work, we assume a thermal background of three degenerate mass neutrino species.

The main results with massive neutrinos are shown in Figure \ref{fig:mnu}  and tabulated in Table \ref{tab:cosmo}. 
Adding the SPT cluster sample to the CMB + H$_0$ + BAO data leads to a small preference for a positive neutrino mass sum. 
If we fit the 1-D posterior on the total neutrino mass with a Gaussian (avoiding the bias to the median and 68\% interval values due to the positivity prior), the preferred value is $\sum m_\nu =0.17 \pm 0.13$\,eV. 
The uncertainties on the neutrino mass tighten with the addition of the cluster data, but the shift in the peak likelihood towards higher masses means that the 95\% confidence upper limit on $\sum m_\nu$ is nearly unchanged:  $\sum m_\nu < \mnunoclus\,$eV without clusters and $\sum m_\nu < \mnu\,$eV with clusters. 
This improvement is largely due to the tighter constraint on $\sigma_8$ derived from the cluster data. 
The \sptcl{} data tightens the $\sigma_8$ measurement from the  \nuset{} data from \sigmaeightmnunoclus{} to \sigmaeightmnu{}.

We again compare these results to those reported by B11. 
B11 report $\sigma_8 = 0.770 \pm 0.026$ and a 95\% CL upper limit of $\sum m_\nu < 0.33\,$eV for the   CMB + H$_0$ + BAO + \sptcl{}(B11) data. 
The median $\sigma_8$ value has shifted down slightly in this work leading to a higher $\sum m_\nu$ limit. 
The parameter uncertainties are essentially unchanged.

\subsection{Prospects for Further Improvement}

The cosmological results in this paper are derived from a high-purity and high-redshift subsample of the catalog consisting of \ncosmo{} galaxy clusters. 
The full SPT survey covers approximately 3.5 times the sky area used in this work and is being used to produce a similar high purity catalog with 
3.5 times as many clusters. 
Realizing the scientific potential of this sample will require significant improvements to the current mass calibration. 
We have simulated the impact of a more accurate mass calibration on both the current catalog and the full SPT survey. 
A 5\% mass calibration would tighten the current constraints  to $\sigma(w) = 0.043$ (for \allset) and $\sigma(\sum m_\nu) = 0.10$\,eV (for \nuset) respectively, a factor of 1.3 better than those listed in Table~\ref{tab:cosmo} for the current mass calibration. 
Determining the mass calibration to better than 5\% would have little impact with the current catalog. 
However, it would significantly improve constraints for the $\times 3.5$~larger, full SPT sample and could make possible a significant detection of the sum of the neutrino masses.

As laid out by B11, four approved observation programs are being pursued by the
SPT collaboration to independently test the cluster mass calibration,
with the goal of reducing this uncertainty to a level $\lesssim$\,5\%. 
First, X-ray observations with \chandra{} are scheduled for the 80 most-significant SPT cluster detections at $z > 0.4$. 
Second,  we have been awarded time for  weak lensing observations of $\sim$35 SPT-detected clusters spanning $0.30 < z < 1.3$ using the  Magellan and \hubble\ telescopes.  
Third, we have been awarded time for optical velocity  
dispersion observations of $\sim$100 SPT-detected clusters using the Very Large Telescope (VLT)  
and a large NOAO program on Gemini South.  
Fourth, the DES will also yield weak lensing mass estimates (S/N $\sim$ 1) of all SPT cluster candidates. 
The combination of the X-ray, velocity dispersion, and weak lensing observations will enable valuable cross-checks between these different mass estimates and
should lead to significant improvements in the cluster mass calibration.

\section{Conclusions}
\label{sec:conclusions}

We have presented a catalog of 224 cluster candidates detected with signal-to-noise greater than 4.5 in 720\,deg$^2$ of the SPT survey. 
Using optical/NIR follow-up data, we have detected clear counterparts 
for \nconfirmfourf\ of these candidates, of which 
\nconfirmnewspt\ were first identified as galaxy clusters in the SPT data. 
The observed purity of the full sample is 71\%; the purity rises to 95\% for the 121 candidates detected at a signal-to-noise greater than 5. 
We report photometric and in some cases spectroscopic redshifts for these 
galaxy clusters, finding redshifts between $0.0552 < z <  1.37$ with a median redshift of $z=\medianz$.
We also estimate the masses based on simulations and X-ray observations, and find the median mass of the sample is $M_{500} =\medianm$. 
This catalog expands the total number of published, optically confirmed galaxy clusters discovered with the SPT to \nconfirmnewallspt{}\ and triples the total number of SZ-identified galaxy clusters. 

We extend the cosmological fitting algorithm for SZ clusters presented by \citetalias{vanderlinde10} and B11 in two ways. 
First, we implement an improved treatment of unconfirmed cluster candidates. 
This improvement has minimal impact given the high purity ($\sim$95\%) of the $\nSN >$5 catalog. 
Second and more importantly, we develop a framework for combining cluster counts from fields observed with 
different noise levels, using simulations to recalibrate the SZ detection significance in each field.
This framework will be essential for optimally analyzing the final SPT catalog. 

We derive cosmological constraints based on the measured cluster abundances.
In these analyses, we limit the cluster sample to the \ncosmo\ cluster candidates detected with signal-to-noise $\nSN > 5$ and $z > 0.3$ (or optically-unconfirmed). 
Using just the information from these clusters and a BBN + $H_0$ prior, 
we see a sizeable improvement to the 
constraints on a \lcdm \ cosmological model compared to the constraints reported
in B11 with a smaller cluster sample and similar priors. 
However, when additional data (${\rm CMB + BAO + SNe}$) are added, the 
constraints from the cluster sample presented here are similar to those from B11.
This is to be expected, because the B11 constraints were already limited by mass 
calibration uncertainty, not cluster sample size.

Adding the SPT cluster data to CMB+BAO+H$_0$+SNe data constrains the equation of state of dark energy to be $w = \wWCDM$. 
The uncertainty is a factor of 1.3 smaller than that without 
the SPT catalog and the preferred value remains consistent with a cosmological constant. 
The addition of SPT cluster data also reduces the uncertainty on $\sigma_8$ in a \wCDM\ cosmology by a factor of 1.4 from $\sigma_8 = 0.840 \pm 0.038$ to $\sigma_8 = \sigmaeightWCDM$.

We also use the measured SZ cluster counts to constrain $\sigma_8$ and the sum of
the neutrino masses.  In an extension to the \lcdm \ model that includes massive neutrinos, the SZ cluster counts tighten the
$\sigma_8$ constraint by a factor of 1.4 when added to the CMB+BAO+H$_0$ data. 
This leads to a small preference for positive neutrino masses  with $\sum m_\nu = 0.17 \pm 0.13$\,eV. 
The 95\% confidence upper limit on the total neutrino mass slightly decreases from \mnunoclus\,eV to \mnu\,eV.  
The relative improvement  to the upper limit is less than would be expected because of the preference for higher neutrino masses.

The SPT survey of 2500\,\sqdeg\ was completed in November, 2011. 
The survey area, comprising 6\% of the total sky, has been mapped to depths of approximately 40, 18, and $70\, \mukcmb$-arcmin at 95, 150, and $220\,$GHz respectively. 
These depths are roughly equal to those of the 2009 data presented here. 
The survey should detect $\sim$\nfullsurvey{} optically-confirmed galaxy clusters at signal-to-noise $\nSN > 4.5$, with a median redshift of $\sim0.5$ and a median mass of $\mass \sim  3 \times 10^{14} M_{\odot} h^{-1}_{70}$.
Ongoing X-ray, weak lensing, and optical velocity dispersion observations of 
SPT SZ-selected clusters will be used to produce an improved cluster mass calibration of the sample.  
The full SPT survey and improved mass calibration will lead to 
constraints on the dark energy equation of state, $w$, better than current constraints from the combination of 
CMB+BAO+SNe data and will provide an independent systematic test of the 
standard dark energy paradigm by measuring the effect of dark energy on the 
growth of structure.  
Furthermore, the combination of CMB+BAO+SNe constraints with those from the full 
SPT cluster sample will break parameter degeneracies that exist in either 
data set alone, resulting in significantly tighter constraints on dark 
energy. The addition of the SPT cluster abundance data is also already leading 
to tighter constraints on the sum of the neutrino masses; with 
the ongoing program to improve the cluster mass calibration, it may be possible to 
produce a significant detection of non-zero neutrino mass.
\acknowledgments

The South Pole Telescope program is supported by the National Science
Foundation through grant ANT-0638937.  Partial support is also
provided by the NSF Physics Frontier Center grant PHY-0114422 to the
Kavli Institute of Cosmological Physics at the University of Chicago,
the Kavli Foundation, and the Gordon and Betty Moore Foundation.  
Galaxy cluster research
at Harvard is supported by NSF grant AST-1009012.  Galaxy cluster
research at SAO is supported in part by NSF grants AST-1009649 and
MRI-0723073.  The McGill group acknowledges funding from the National
Sciences and Engineering Research Council of Canada, Canada Research
Chairs program, and the Canadian Institute for Advanced Research.
X-ray research at the CfA is supported through NASA Contract NAS
8-03060.  This work is based in part on observations made with the Spitzer Space Telescope, which is operated by the Jet Propulsion Laboratory, California Institute of Technology under a contract with NASA. Support for this work was provided by NASA through an award issued by JPL/Caltech. The Munich group acknowledges support from the Excellence
Cluster Universe and the DFG research program TR33.  
 R.J.F.\ is supported by a Clay Fellowship.  B.A.B\ is supported by a KICP
Fellowship, M.Bautz acknowledges support from contract
2834-MIT-SAO-4018 from the Pennsylvania State University to the
Massachusetts Institute of Technology.  M.D.\ acknowledges support
from an Alfred P.\ Sloan Research Fellowship, W.F.\ and C.J.\
acknowledge support from the Smithsonian Institution, and B.S.\
acknowledges support from the Brinson Foundation.

Support for X-ray analysis was provided by NASA through Chandra
Award Numbers 12800071, 12800088, and G02-13006A issued by the Chandra X-ray
Observatory Center, which is operated by the Smithsonian Astrophysical
Observatory for and on behalf of NASA under contract NAS8-03060.
Optical imaging data from the Blanco 4~m at Cerro Tololo Interamerican
Observatories (programs 2005B-0043, 2009B-0400, 2010A-0441,
2010B-0598) and spectroscopic observations from VLT programs
086.A-0741 and 286.A-5021 and Gemini program GS-2009B-Q-16 were
included in this work.  Additional data were obtained with the 6.5~m
Magellan Telescopes located at the Las Campanas Observatory,
Chile. 

We acknowledge the use of the Legacy Archive for
Microwave Background Data Analysis (LAMBDA).  Support for LAMBDA is
provided by the NASA Office of Space Science.  
This research used resources of the National Energy Research Scientific Computing Center, which is supported by the Office of Science of the U.S. Department of Energy under Contract No. DE-AC02-05CH11231. 
This research has made use of the SIMBAD database,
operated at CDS, Strasbourg, France, and the NASA/IPAC Extragalactic Database (NED) 
which is operated by the Jet Propulsion Laboratory, California Institute of Technology, 
under contract with the National Aeronautics and Space Administration.

{\it Facilities:}
\facility{Blanco (MOSAIC, NEWFIRM)},
\facility{CXO (ACIS)},
\facility{Gemini-S (GMOS)},
\facility{Magellan:Baade (IMACS)},
\facility{Magellan:Clay (LDSS3)},
\facility{Spitzer (IRAC)},
\facility{South Pole Telescope},
\facility{Swope (CCD)},
\facility{XMM-Newton (EPIC)}

\bibliography{../../BIBTEX/spt}

\appendix

\clearpage
\LongTables 
\pagestyle{empty}
\begin{center}
\def\arraystretch{1.2}
\tabletypesize{\scriptsize}
\begin{deluxetable}{l cc | c | ccc  | cc   | c | c }
\tablecaption{Galaxy clusters above $4.5 \sigma$ in $720$ square degrees observed by the SPT.
\label{tab:catalog}}
\tablehead{
\multicolumn{3}{c}{{\bf ID \& coordinates:}} & 
\multicolumn{1}{c}{{\bf Y$_{\rm SZ}$} $\times 10^6$} &
\multicolumn{3}{c}{\bf Significances} &
\multicolumn{2}{c}{\bf Best} &
\multicolumn{1}{c}{\bf Redshift} &
\multicolumn{1}{c}{\bf M$_{500}$}  \\
\colhead{SPT ID} & 
\colhead{RA} & 
\colhead{DEC}  & 
\colhead{(${\rm arcmin}^2$)}  & 
\colhead{$\theta_c$=0.5'} & 
\colhead{1.5'} & 
\colhead{2.5'} & 
\colhead{\nSN{}} & 
\colhead{$\theta_c$} & 
\colhead{} &
\colhead{($10^{14} h_{70}^{-1} M_\odot$)} \\
}

\startdata 
\input{catalog_data}
\tablenotetext{A}{Unconfirmed cluster candidate which is either above the quoted redshift limit or a false detection.}
\tablecomments{Galaxy cluster candidates selected above a significance of 4.5 in the first $720\,$\sqdeg{} of the SPT survey. 
Galaxy clusters marked by a `*' have X-ray data that are used in the cosmological analysis (see A11 and B11 for a description of the X-ray data).
For each candidate, we report the detection significance of each candidate in the `Best' column, as well as the significances at a fixed set of three core radii (\S\ref{sec:extract}). 
We also report the position and (if confirmed) redshift (\S\ref{sec:optical}). 
Spectroscopic redshifts are quoted without uncertainties. 
The integrated Y$_{\rm SZ}$ is reported for a 1$^\prime$ aperture (\S\ref{subsec:yint}). 
Finally, we report a mass estimate for each confirmed cluster marginalized over the \lcdm{} chain (\S\ref{sec:mass}).
}
\end{deluxetable}
\end{center}
\clearpage

\pagestyle{plain}

\end{document}

%% file: tab_xray.tex
 SPT-CL J0233-5819 & 1RXS J023303.1-581939 &  13 & $ 0.6630 $ & $   0.0295 \pm   0.0131 $ &     2.90 &     4.63  \\  
 SPT-CL J0234-5831 & 1RXS J023443.1-583114 &   4 & $ 0.4150 $ & $   0.0800 \pm   0.0200 $ &     7.95 &     4.12  \\  
 SPT-CL J0254-5857 & 1RXS J025427.2-585736 &  80 & $ 0.4380 $ & $   0.0846 \pm   0.0305 $ &     7.54 &     4.41  \\  
 SPT-CL J0257-5842 & 1RXS J025744.7-584120 & 116 & $ 0.43 \pm 0.03  $ & $   0.0725 \pm   0.0298 $ &     6.10 &     3.66  \\  
 SPT-CL J0324-6236 & 1RXS J032412.7-623553 &  13 & $ 0.72 \pm 0.04  $ & $   0.0260 \pm   0.0121 $ &     2.56 &     4.85  \\  
 SPT-CL J0328-5541 & 1RXS J032833.5-554232 &  68 & $ 0.0844 $ & $   0.5700 \pm   0.0300 $ &    47.28 &     1.31  \\  
 SPT-CL J0333-5842 & 1RXS J033317.3-584244 &  38 & $ 0.47 \pm 0.03  $ & $   0.0125 \pm   0.0056 $ &     1.13 &     0.82  \\  
 SPT-CL J0337-6300 & 1RXS J033754.5-630122 &  49 & $ 0.45 \pm 0.03  $ & $   0.0166 \pm   0.0079 $ &     1.78 &     1.22  \\  
 SPT-CL J0343-5518 & 1RXS J034259.3-551905 &  58 & $ 0.51 \pm 0.03  $ & $   0.0167 \pm   0.0071 $ &     1.35 &     1.21  \\  
 SPT-CL J0354-5904 & 1RXS J035420.7-590545 &  92 & $ 0.46 \pm 0.03  $ & $   0.0105 \pm   0.0049 $ &     0.91 &     0.62  \\  
 SPT-CL J0402-6129 & 1RXS J040245.7-612939 &  32 & $ 0.52 \pm 0.03  $ & $   0.0082 \pm   0.0039 $ &     0.74 &     0.70  \\  
 SPT-CL J0403-5719 & 1RXS J040352.3-571936 &  10 & $ 0.46 \pm 0.03  $ & $   0.0391 \pm   0.0081 $ &     3.06 &     2.11  \\  
 SPT-CL J0404-6510 & 1RXS J040421.6-651004 &  72 & $ 0.14 \pm 0.02  $ & $   0.1300 \pm   0.0200 $ &    13.55 &     0.74  \\  
 SPT-CL J0410-6343 & 1RXS J041009.3-634319 &  43 & $ 0.50 \pm 0.03  $ & $   0.0291 \pm   0.0103 $ &     2.88 &     2.44  \\  
 SPT-CL J0411-6340 & 1RXS J041129.7-634133 &  47 & $ 0.14 \pm 0.02  $ & $   0.2600 \pm   0.0300 $ &    25.83 &     1.26  \\  
 SPT-CL J0412-5743 & 1RXS J041206.3-574313 &   3 & $ 0.39 \pm 0.03  $ & $   0.0231 \pm   0.0074 $ &     1.85 &     0.88  \\  
 SPT-CL J0423-5506 & 1RXS J042315.7-550710 &  58 & $ 0.20 \pm 0.03  $ & $   0.0332 \pm   0.0125 $ &     2.24 &     0.25  \\  
 SPT-CL J0431-6126 & 1RXS J043126.6-612622 &  40 & $ 0.0577 $ & $   0.9800 \pm   0.0700 $ &    82.50 &     1.14  \\  
 SPT-CL J0509-5342* & 1RXS J050921.2-534159 &  18 & $ 0.4626 $ & $   0.0351 \pm   0.0118 $ &     2.79 &     1.94  \\  
 SPT-CL J0516-5430 & 1RXS J051634.0-543104 &  44 & $ 0.2950 $ & $   0.1200 \pm   0.0200 $ &    10.86 &     2.71  \\  
 SPT-CL J0521-5104 & 1RXS J052113.2-510419 &  37 & $ 0.6755 $ & $   0.0135 \pm   0.0062 $ &     1.20 &     2.04  \\  
 SPT-CL J0539-5744 & 1RXS J054010.1-574354 &  91 & $ 0.76 \pm 0.03  $ & $   0.0123 \pm   0.0053 $ &     1.47 &     3.32  \\  
 SPT-CL J0546-5345* & 1RXS J054638.7-534434 &  69 & $ 1.0670 $ & $   0.0123 \pm   0.0044 $ &     1.59 &     7.55  \\  
 SPT-CL J0551-5709* & 1RXS J055126.4-570843 &  91 & $ 0.4230 $ & $   0.0271 \pm   0.0053 $ &     3.41 &     1.96  \\  
 SPT-CL J0559-5249* & 1RXS J055942.1-524950 &  15 & $ 0.6112 $ & $   0.0109 \pm   0.0042 $ &     1.29 &     1.65  \\  
 SPT-CL J2011-5725 & 1RXS J201127.9-572507 &  28 & $ 0.2786 $ & $   0.1100 \pm   0.0300 $ &    12.25 &     2.80  \\  
 SPT-CL J2012-5649 & 1RXS J201238.3-565038 & 103 & $ 0.0552 $ & $   1.1400 \pm   0.0900 $ &   130.40 &     0.96  \\  
 SPT-CL J2016-4954 & 1RXS J201603.5-495530 &  47 & $ 0.26 \pm 0.03  $ & $   0.0273 \pm   0.0127 $ &     2.98 &     0.59  \\  
 SPT-CL J2018-4528 & 1RXS J201828.7-452720 &  95 & $ 0.41 \pm 0.03  $ & $   0.0298 \pm   0.0129 $ &     2.97 &     1.62  \\  
 SPT-CL J2021-5256 & 1RXS J202155.7-525721 &  52 & $ 0.11 \pm 0.02  $ & $   0.0600 \pm   0.0200 $ &     6.62 &     0.20  \\  
 SPT-CL J2023-5535 & 1RXS J202321.2-553534 &   9 & $ 0.2320 $ & $   0.0900 \pm   0.0200 $ &    10.58 &     1.54  \\  
 SPT-CL J2025-5117 & 1RXS J202554.4-511647 &  41 & $ 0.18 \pm 0.02  $ & $   0.0500 \pm   0.0100 $ &     5.12 &     0.43  \\  
 SPT-CL J2032-5627 & 1RXS J203215.2-562753 &  47 & $ 0.2840 $ & $   0.0542 \pm   0.0180 $ &     6.64 &     1.54  \\  
 SPT-CL J2121-6335 & 1RXS J212157.9-633459 & 103 & $ 0.23 \pm 0.03  $ & $   0.1000 \pm   0.0200 $ &     9.85 &     1.48  \\  
 SPT-CL J2130-6458 & 1RXS J213056.1-645909 &  36 & $ 0.3160 $ & $   0.0437 \pm   0.0189 $ &     4.33 &     1.28  \\  
 SPT-CL J2136-4704 & 1RXS J213624.5-470453 &  38 & $ 0.4250 $ & $   0.0286 \pm   0.0114 $ &     2.58 &     1.50  \\  
 SPT-CL J2138-6007 & 1RXS J213801.2-600801 &   5 & $ 0.3190 $ & $   0.0750 \pm   0.0211 $ &     7.74 &     2.26  \\  
 SPT-CL J2145-5644 & 1RXS J214559.3-564455 &  55 & $ 0.4800 $ & $   0.0413 \pm   0.0162 $ &     4.01 &     2.91  \\  
 SPT-CL J2146-5736 & 1RXS J214643.9-573723 &  43 & $ 0.60 \pm 0.03  $ & $   0.0277 \pm   0.0119 $ &     2.64 &     3.36  \\  
 SPT-CL J2201-5956 & 1RXS J220157.8-595648 &  33 & $ 0.0983 $ & $   1.0800 \pm   0.0400 $ &   108.10 &     2.57  \\  
 SPT-CL J2259-5432 & 1RXS J225957.0-543118 &  51 & $ 0.44 \pm 0.04  $ & $   0.0225 \pm   0.0098 $ &     1.68 &     1.07  \\  
 SPT-CL J2259-5617 & 1RXS J230001.2-561709 &  17 & $ 0.17 \pm 0.02  $ & $   0.1400 \pm   0.0200 $ &    11.29 &     0.87  \\  
 SPT-CL J2300-5331 & 1RXS J230039.8-533118 &  28 & $ 0.2620 $ & $   0.0800 \pm   0.0200 $ &     5.81 &     1.16  \\  
 SPT-CL J2332-5358* & 1RXS J233224.3-535840 &  17 & $ 0.4020 $ & $   0.1600 \pm   0.0300 $ &    12.29 &     6.23  \\  
 SPT-CL J2337-5942* & 1RXS J233726.6-594205 &  18 & $ 0.7750 $ & $   0.0271 \pm   0.0136 $ &     2.18 &     4.64  \\  

%% file: tab_other.tex
SPT-CL J0254-5857 & SPT-CL J0254-5856, A & A,B & $  0.4380$ & $  0.4380$, A \\ 
SPT-CL J0328-5541 & ACO 3126, C & B,C,D,E,F,G & $ 0.0844$ & $  0.0844$, H \\ 
SPT-CL J0404-6510 & ACO 3216, C & C & $ 0.14 \pm  0.02$ & $  0.14$, I \\ 
SPT-CL J0411-6340 & ACO 3230, C & C & $ 0.14 \pm  0.02$ & $  0.14$, I \\ 
SPT-CL J0431-6126 & Ser 40-6, J & B,C,D,E,F,G,J,K & $ 0.0577$ & $  0.0577$, H \\ 
SPT-CL J0458-5741 & ACO 3298, C & C & Unconfirmed & - \\ 
SPT-CL J0509-5342 & SPT-CL 0509-5342, L & L,M,N & $  0.4626$ & $  0.4626$, O \\ 
SPT-CL J0511-5154 & SCSO J051145-515430, P & M,P & $  0.6450$ & $  0.74$, O \\ 
SPT-CL J0516-5430 & ACO S 0520, C & B,C,G,L,M,N,P & $  0.2950$ & $  0.2950$, G \\ 
SPT-CL J0521-5104 & SCSO J052113-510418, P & M,P & $  0.6755$ & $  0.72$, O \\ 
SPT-CL J0522-5026 & SCSO J052200-502700, P & P & $ 0.53 \pm  0.04$ & $0.50$, P \\ 
SPT-CL J0528-5300 & SPT-CL 0528-5300, L & L,M,N,P & $  0.7648$ & $  0.7648$, O \\ 
SPT-CL J0546-5345 & SPT-CL 0547-5345, L & L,M,N & $  1.0670$ & $  1.0670$, Q \\ 
SPT-CL J0559-5249 & SPT-CL J0559-5249, M & M,N & $  0.6112$ & $  0.6112$, O \\ 
SPT-CL J2011-5725 & RXC J2011.3-5725, G & G & $  0.2786$ & $  0.2786$, G \\ 
SPT-CL J2012-5649 & Str 2008-569, K & B,C,E,G,K,R & $  0.0552$ & $  0.0552$, H \\ 
SPT-CL J2020-4646 & ACO 3673, C & C & $ 0.19 \pm  0.02$ & - \\ 
SPT-CL J2021-5256 & Ser 138-5, J & C,G,J & $ 0.11 \pm  0.02$ & - \\ 
SPT-CL J2023-5535 & RXC J2023.4-5535, G & A,B,G & $  0.2320$ & $  0.2320$, G \\ 
SPT-CL J2025-5117 & ACO S 0871, C & C & $ 0.18 \pm  0.02$ & - \\ 
SPT-CL J2032-5627 & ClG 2028.3-5637, S & C,G,S & $  0.2840$ & $  0.0608$, H \\ 
SPT-CL J2055-5456 & ACO 3718, C & C,G & $ 0.13 \pm  0.02$ & - \\ 
SPT-CL J2059-5018 & ACO S 0912, C & C & $ 0.41 \pm  0.03$ & - \\ 
SPT-CL J2101-5542 & ACO 3732, C & C & $ 0.20 \pm  0.02$ & - \\ 
SPT-CL J2121-6335 & ACO S 0937, C & C & $ 0.23 \pm  0.03$ & - \\ 
SPT-CL J2201-5956 & ClG 2158.3-6011, S & A,B,C,D,E,F,G,S & $  0.0972$ & $  0.0972$, H \\ 
SPT-CL J2259-5617 & ACO 3950, C & C,M & $ 0.17 \pm  0.02$ & - \\ 
SPT-CL J2300-5331 & ACO S 1079, C & C,M & $  0.2620$ & $  0.29$, O \\ 
SPT-CL J2332-5358 & SCSO J233227-535827, P & M,P & $  0.4020$ & $  0.32$, T \\ 
SPT-CL J2351-5452 & SCSO J235138-545253, P & P & $  0.3838$ & $  0.3838$, U \\ 

%% file: tab_missed.tex
SPT-CL J0334-6008 &      53.7116 &     -60.1541 &     6.97 &     1.25 \\ 
SPT-CL J0434-5727 &      68.6517 &     -57.4568 &     5.07 &     0.75 \\ 
SPT-CL J0442-5905 &      70.6707 &     -59.0975 &     6.42 &     0.25 \\ 
SPT-CL J2142-6419 &     325.7280 &     -64.3268 &    11.01 &     0.25 \\ 
SPT-CL J2154-5952 &     328.7210 &     -59.8878 &     7.16 &     0.50 \\ 
SPT-CL J2154-5936 &     328.7230 &     -59.6121 &     6.28 &     0.50 \\ 

%% file: catalog_data.tex
SPT-CL J0000-5748* & $    0.2496 $  & $  -57.8066 $  & $ 107 \pm 24 $ & $   5.48 $  & $   4.84 $  & $   4.38 $  & $   5.48 $  & $ 0.50 $  & $ 0.7019 $ & $   4.32 \pm   0.75 $ \\
SPT-CL J0201-6051 & $   30.3933 $  & $  -60.8592 $  & $ 73 \pm 19 $ & $   4.44 $  & $   3.50 $  & $   2.39 $  & $   4.83 $  & $ 0.25 $  & $> 1.05^A  $ & -\\
SPT-CL J0203-5651 & $   30.8309 $  & $  -56.8612 $  & $ 74 \pm 19 $ & $   4.78 $  & $   4.02 $  & $   2.45 $  & $   4.98 $  & $ 0.25 $  & $> 1.00^A  $ & -\\
SPT-CL J0205-5829 & $   31.4437 $  & $  -58.4856 $  & $ 185 \pm 19 $ & $  10.39 $  & $   8.96 $  & $   7.19 $  & $  10.54 $  & $ 0.25 $  & $ 1.3200 $ & $   4.82 \pm   0.96 $ \\
SPT-CL J0205-6432 & $   31.2786 $  & $  -64.5461 $  & $ 103 \pm 19 $ & $   6.02 $  & $   5.04 $  & $   3.91 $  & $   6.02 $  & $ 0.50 $  & $ 0.7440 $ & $   3.39 \pm   0.82 $ \\
SPT-CL J0209-5452 & $   32.3491 $  & $  -54.8794 $  & $ 83 \pm 21 $ & $   4.52 $  & $   4.40 $  & $   4.05 $  & $   4.52 $  & $ 0.50 $  & $ 0.41 \pm 0.04  $ & $   2.57 \pm   0.86 $ \\
SPT-CL J0211-5712 & $   32.8232 $  & $  -57.2157 $  & $ 66 \pm 21 $ & $   4.65 $  & $   3.66 $  & $   3.05 $  & $   4.77 $  & $ 0.25 $  & $> 1.03^A  $ & -\\
SPT-CL J0216-5730 & $   34.1363 $  & $  -57.5100 $  & $ 85 \pm 20 $ & $   4.72 $  & $   4.48 $  & $   3.79 $  & $   4.72 $  & $ 0.50 $  & $> 1.03^A  $ & -\\
SPT-CL J0216-6409 & $   34.1723 $  & $  -64.1562 $  & $ 94 \pm 19 $ & $   5.53 $  & $   4.92 $  & $   4.27 $  & $   5.54 $  & $ 0.25 $  & $ 0.62 \pm 0.04  $ & $   3.18 \pm   0.82 $ \\
SPT-CL J0218-5826 & $   34.6251 $  & $  -58.4386 $  & $ 78 \pm 20 $ & $   4.48 $  & $   4.15 $  & $   3.17 $  & $   4.54 $  & $ 0.25 $  & $ 0.56 \pm 0.04  $ & $   2.50 \pm   0.86 $ \\
SPT-CL J0221-6212 & $   35.4382 $  & $  -62.2044 $  & $ 65 \pm 19 $ & $   4.53 $  & $   3.66 $  & $   2.96 $  & $   4.71 $  & $ 0.25 $  & $> 1.20^A  $ & -\\
SPT-CL J0230-6028 & $   37.6410 $  & $  -60.4694 $  & $ 98 \pm 19 $ & $   5.79 $  & $   4.87 $  & $   3.84 $  & $   5.88 $  & $ 0.25 $  & $ 0.74 \pm 0.09  $ & $   3.29 \pm   0.79 $ \\
SPT-CL J0233-5819 & $   38.2561 $  & $  -58.3269 $  & $ 131 \pm 20 $ & $   6.34 $  & $   6.44 $  & $   5.67 $  & $   6.64 $  & $ 1.25 $  & $ 0.6630 $ & $   3.79 \pm   0.86 $ \\
SPT-CL J0234-5831 & $   38.6790 $  & $  -58.5217 $  & $ 270 \pm 19 $ & $  14.65 $  & $  13.03 $  & $  10.54 $  & $  14.65 $  & $ 0.50 $  & $ 0.4150 $ & $   7.71 \pm   1.50 $ \\
SPT-CL J0239-6148 & $   39.9120 $  & $  -61.8032 $  & $ 74 \pm 19 $ & $   4.50 $  & $   3.46 $  & $   2.75 $  & $   4.67 $  & $ 0.25 $  & $> 1.06^A  $ & -\\
SPT-CL J0240-5946 & $   40.1620 $  & $  -59.7703 $  & $ 169 \pm 19 $ & $   8.99 $  & $   8.64 $  & $   7.63 $  & $   9.04 $  & $ 0.75 $  & $ 0.4000 $ & $   5.32 \pm   1.11 $ \\
SPT-CL J0240-5952 & $   40.1982 $  & $  -59.8785 $  & $ 74 \pm 19 $ & $   4.53 $  & $   3.64 $  & $   2.50 $  & $   4.65 $  & $ 0.25 $  & $ 0.62 \pm 0.03  $ & $   2.54 \pm   0.82 $ \\
SPT-CL J0242-6039 & $   40.6551 $  & $  -60.6526 $  & $ 90 \pm 20 $ & $   4.87 $  & $   4.77 $  & $   4.26 $  & $   4.92 $  & $ 1.00 $  & $> 1.50^A  $ & -\\
SPT-CL J0243-5930 & $   40.8616 $  & $  -59.5132 $  & $ 126 \pm 19 $ & $   7.30 $  & $   6.23 $  & $   5.18 $  & $   7.42 $  & $ 0.25 $  & $ 0.65 \pm 0.04  $ & $   4.25 \pm   0.89 $ \\
SPT-CL J0249-5658 & $   42.4068 $  & $  -56.9764 $  & $ 96 \pm 20 $ & $   5.37 $  & $   5.21 $  & $   4.48 $  & $   5.44 $  & $ 0.75 $  & $ 0.22 \pm 0.02  $ & $   3.43 \pm   0.93 $ \\
SPT-CL J0253-6046 & $   43.4605 $  & $  -60.7744 $  & $ 86 \pm 21 $ & $   4.61 $  & $   4.73 $  & $   4.27 $  & $   4.83 $  & $ 1.25 $  & $ 0.46 \pm 0.02  $ & $   2.79 \pm   0.86 $ \\
SPT-CL J0254-5857 & $   43.5729 $  & $  -58.9526 $  & $ 295 \pm 22 $ & $  13.61 $  & $  14.42 $  & $  13.82 $  & $  14.42 $  & $ 1.50 $  & $ 0.4380 $ & $   7.61 \pm   1.46 $ \\
SPT-CL J0254-6051 & $   43.6015 $  & $  -60.8643 $  & $ 127 \pm 21 $ & $   6.35 $  & $   6.59 $  & $   5.75 $  & $   6.71 $  & $ 1.00 $  & $ 0.46 \pm 0.02  $ & $   4.04 \pm   0.89 $ \\
SPT-CL J0256-5617 & $   44.1009 $  & $  -56.2973 $  & $ 136 \pm 19 $ & $   7.54 $  & $   6.98 $  & $   5.83 $  & $   7.54 $  & $ 0.50 $  & $ 0.64 \pm 0.04  $ & $   4.32 \pm   0.89 $ \\
SPT-CL J0257-5732 & $   44.3516 $  & $  -57.5423 $  & $ 95 \pm 19 $ & $   5.28 $  & $   4.53 $  & $   3.51 $  & $   5.40 $  & $ 0.25 $  & $ 0.4340 $ & $   3.21 \pm   0.86 $ \\
SPT-CL J0257-5842 & $   44.3924 $  & $  -58.7116 $  & $ 105 \pm 21 $ & $   5.35 $  & $   5.19 $  & $   4.71 $  & $   5.38 $  & $ 1.00 $  & $ 0.43 \pm 0.03  $ & $   3.21 \pm   0.86 $ \\
SPT-CL J0257-6050 & $   44.3354 $  & $  -60.8450 $  & $ 96 \pm 23 $ & $   4.46 $  & $   4.68 $  & $   4.32 $  & $   4.76 $  & $ 1.25 $  & $ 0.50 \pm 0.04  $ & $   2.71 \pm   0.86 $ \\
SPT-CL J0258-5756 & $   44.5562 $  & $  -57.9438 $  & $ 95 \pm 22 $ & $   4.06 $  & $   4.50 $  & $   4.18 $  & $   4.50 $  & $ 1.50 $  & $> 1.05^A  $ & -\\
SPT-CL J0300-6315 & $   45.1430 $  & $  -63.2643 $  & $ 73 \pm 22 $ & $   3.61 $  & $   4.29 $  & $   4.85 $  & $   4.88 $  & $ 2.75 $  & $> 1.50^A  $ & -\\
SPT-CL J0301-6456 & $   45.4780 $  & $  -64.9470 $  & $ 78 \pm 20 $ & $   4.79 $  & $   4.07 $  & $   2.99 $  & $   4.94 $  & $ 0.25 $  & $ 0.66 \pm 0.03  $ & $   2.68 \pm   0.82 $ \\
SPT-CL J0307-6226 & $   46.8335 $  & $  -62.4336 $  & $ 156 \pm 20 $ & $   8.15 $  & $   8.09 $  & $   6.94 $  & $   8.32 $  & $ 0.75 $  & $ 0.59 \pm 0.04  $ & $   4.75 \pm   0.96 $ \\
SPT-CL J0311-6354 & $   47.8283 $  & $  -63.9083 $  & $ 136 \pm 20 $ & $   7.06 $  & $   7.11 $  & $   6.29 $  & $   7.33 $  & $ 1.00 $  & $ 0.30 \pm 0.02  $ & $   4.54 \pm   0.96 $ \\
SPT-CL J0313-5645 & $   48.2604 $  & $  -56.7554 $  & $ 90 \pm 20 $ & $   4.70 $  & $   4.50 $  & $   3.82 $  & $   4.82 $  & $ 0.75 $  & $ 0.61 \pm 0.04  $ & $   2.68 \pm   0.82 $ \\
SPT-CL J0316-6059 & $   49.2179 $  & $  -60.9849 $  & $ 92 \pm 19 $ & $   4.45 $  & $   3.95 $  & $   3.28 $  & $   4.59 $  & $ 0.25 $  & $> 1.50^A  $ & -\\
SPT-CL J0317-5935 & $   49.3208 $  & $  -59.5856 $  & $ 100 \pm 19 $ & $   5.80 $  & $   5.00 $  & $   4.00 $  & $   5.91 $  & $ 0.25 $  & $ 0.4690 $ & $   3.54 \pm   0.89 $ \\
SPT-CL J0320-5800 & $   50.0316 $  & $  -58.0084 $  & $ 74 \pm 21 $ & $   3.93 $  & $   4.47 $  & $   4.42 $  & $   4.54 $  & $ 2.25 $  & $> 0.99^A  $ & -\\
SPT-CL J0324-6236 & $   51.0530 $  & $  -62.6018 $  & $ 163 \pm 19 $ & $   8.59 $  & $   8.01 $  & $   6.82 $  & $   8.59 $  & $ 0.50 $  & $ 0.72 \pm 0.04  $ & $   4.75 \pm   0.96 $ \\
SPT-CL J0328-5541 & $   52.1663 $  & $  -55.6975 $  & $ 151 \pm 23 $ & $   6.71 $  & $   6.98 $  & $   6.75 $  & $   7.08 $  & $ 1.75 $  & $ 0.0844 $ & $   4.57 \pm   1.00 $ \\
SPT-CL J0333-5842 & $   53.3195 $  & $  -58.7019 $  & $ 81 \pm 21 $ & $   4.43 $  & $   4.54 $  & $   4.31 $  & $   4.54 $  & $ 1.50 $  & $ 0.47 \pm 0.03  $ & $   2.57 \pm   0.86 $ \\
SPT-CL J0337-6207 & $   54.4720 $  & $  -62.1176 $  & $ 89 \pm 21 $ & $   4.32 $  & $   4.84 $  & $   4.75 $  & $   4.88 $  & $ 1.75 $  & $> 1.28^A  $ & -\\
SPT-CL J0337-6300 & $   54.4685 $  & $  -63.0098 $  & $ 84 \pm 20 $ & $   5.27 $  & $   4.80 $  & $   4.34 $  & $   5.29 $  & $ 0.25 $  & $ 0.45 \pm 0.03  $ & $   3.14 \pm   0.86 $ \\
SPT-CL J0341-5731 & $   55.3979 $  & $  -57.5233 $  & $ 95 \pm 19 $ & $   5.33 $  & $   4.57 $  & $   3.45 $  & $   5.35 $  & $ 0.25 $  & $ 0.62 \pm 0.03  $ & $   3.04 \pm   0.82 $ \\
SPT-CL J0341-6143 & $   55.3485 $  & $  -61.7192 $  & $ 119 \pm 25 $ & $   4.22 $  & $   5.01 $  & $   5.34 $  & $   5.60 $  & $ 3.00 $  & $ 0.64 \pm 0.03  $ & $   3.18 \pm   0.82 $ \\
SPT-CL J0343-5518 & $   55.7634 $  & $  -55.3049 $  & $ 104 \pm 19 $ & $   5.82 $  & $   4.88 $  & $   3.47 $  & $   5.98 $  & $ 0.25 $  & $ 0.51 \pm 0.03  $ & $   3.57 \pm   0.86 $ \\
SPT-CL J0344-5452 & $   56.0926 $  & $  -54.8725 $  & $ 96 \pm 19 $ & $   5.30 $  & $   4.25 $  & $   3.08 $  & $   5.41 $  & $ 0.25 $  & $ 1.01 \pm 0.10  $ & $   2.75 \pm   0.75 $ \\
SPT-CL J0344-5518 & $   56.2101 $  & $  -55.3037 $  & $ 94 \pm 24 $ & $   4.95 $  & $   4.99 $  & $   4.85 $  & $   5.02 $  & $ 1.75 $  & $ 0.37 \pm 0.03  $ & $   3.00 \pm   0.86 $ \\
SPT-CL J0345-6419 & $   56.2518 $  & $  -64.3326 $  & $ 93 \pm 19 $ & $   5.50 $  & $   4.95 $  & $   4.04 $  & $   5.57 $  & $ 0.25 $  & $ 0.93 \pm 0.10  $ & $   2.93 \pm   0.79 $ \\
SPT-CL J0346-5839 & $   56.5745 $  & $  -58.6535 $  & $ 87 \pm 20 $ & $   4.95 $  & $   4.44 $  & $   3.50 $  & $   4.96 $  & $ 0.25 $  & $ 0.74 \pm 0.09  $ & $   2.68 \pm   0.82 $ \\
SPT-CL J0351-5636 & $   57.9312 $  & $  -56.6099 $  & $ 88 \pm 21 $ & $   4.44 $  & $   4.52 $  & $   4.24 $  & $   4.65 $  & $ 0.75 $  & $ 0.39 \pm 0.03  $ & $   2.71 \pm   0.86 $ \\
SPT-CL J0351-5944 & $   57.8654 $  & $  -59.7457 $  & $ 76 \pm 21 $ & $   4.46 $  & $   4.73 $  & $   4.38 $  & $   4.61 $  & $ 1.75 $  & $> 0.99^A  $ & -\\
SPT-CL J0352-5647 & $   58.2366 $  & $  -56.7992 $  & $ 127 \pm 20 $ & $   7.02 $  & $   6.97 $  & $   6.25 $  & $   7.11 $  & $ 0.75 $  & $ 0.66 \pm 0.04  $ & $   4.07 \pm   0.86 $ \\
SPT-CL J0354-5904 & $   58.5611 $  & $  -59.0740 $  & $ 133 \pm 22 $ & $   6.20 $  & $   6.30 $  & $   5.72 $  & $   6.49 $  & $ 1.25 $  & $ 0.46 \pm 0.03  $ & $   3.89 \pm   0.89 $ \\
SPT-CL J0354-6032 & $   58.6744 $  & $  -60.5386 $  & $ 68 \pm 19 $ & $   4.38 $  & $   3.13 $  & $   2.25 $  & $   4.57 $  & $ 0.25 $  & $ 1.07 \pm 0.11  $ & $   2.11 \pm   0.75 $ \\
SPT-CL J0402-6129 & $   60.7066 $  & $  -61.4988 $  & $ 96 \pm 22 $ & $   4.81 $  & $   4.76 $  & $   4.33 $  & $   4.83 $  & $ 1.00 $  & $ 0.52 \pm 0.03  $ & $   2.75 \pm   0.82 $ \\
SPT-CL J0403-5534 & $   60.9479 $  & $  -55.5829 $  & $ 92 \pm 21 $ & $   4.44 $  & $   4.78 $  & $   4.67 $  & $   4.88 $  & $ 1.75 $  & $> 1.50^A  $ & -\\
SPT-CL J0403-5719 & $   60.9670 $  & $  -57.3241 $  & $ 98 \pm 19 $ & $   5.71 $  & $   5.07 $  & $   4.18 $  & $   5.75 $  & $ 0.25 $  & $ 0.46 \pm 0.03  $ & $   3.43 \pm   0.86 $ \\
SPT-CL J0404-6510 & $   61.0556 $  & $  -65.1817 $  & $ 113 \pm 29 $ & $   4.28 $  & $   4.58 $  & $   4.59 $  & $   4.75 $  & $ 2.25 $  & $ 0.14 \pm 0.02  $ & $   2.96 \pm   0.89 $ \\
SPT-CL J0406-5455 & $   61.6922 $  & $  -54.9205 $  & $ 100 \pm 21 $ & $   5.77 $  & $   4.85 $  & $   4.04 $  & $   5.82 $  & $ 0.25 $  & $ 0.74 \pm 0.04  $ & $   3.25 \pm   0.82 $ \\
SPT-CL J0410-5454 & $   62.6154 $  & $  -54.9016 $  & $ 87 \pm 21 $ & $   4.93 $  & $   4.17 $  & $   3.24 $  & $   5.06 $  & $ 0.25 $  & $> 0.98^A  $ & -\\
SPT-CL J0410-6343 & $   62.5158 $  & $  -63.7285 $  & $ 101 \pm 19 $ & $   5.79 $  & $   5.27 $  & $   4.31 $  & $   5.79 $  & $ 0.50 $  & $ 0.50 \pm 0.03  $ & $   3.43 \pm   0.86 $ \\
SPT-CL J0411-5751 & $   62.8432 $  & $  -57.8636 $  & $ 95 \pm 21 $ & $   4.71 $  & $   5.04 $  & $   4.69 $  & $   5.16 $  & $ 1.25 $  & $ 0.77 \pm 0.03  $ & $   2.79 \pm   0.79 $ \\
SPT-CL J0411-6340 & $   62.8597 $  & $  -63.6810 $  & $ 106 \pm 19 $ & $   6.28 $  & $   5.63 $  & $   4.80 $  & $   6.41 $  & $ 0.25 $  & $ 0.14 \pm 0.02  $ & $   4.11 \pm   0.96 $ \\
SPT-CL J0412-5743 & $   63.0245 $  & $  -57.7203 $  & $ 98 \pm 21 $ & $   5.11 $  & $   5.24 $  & $   4.87 $  & $   5.29 $  & $ 1.25 $  & $ 0.39 \pm 0.03  $ & $   3.18 \pm   0.89 $ \\
SPT-CL J0416-6359 & $   64.1618 $  & $  -63.9964 $  & $ 107 \pm 20 $ & $   6.03 $  & $   5.63 $  & $   4.85 $  & $   6.06 $  & $ 0.75 $  & $ 0.28 \pm 0.02  $ & $   3.79 \pm   0.93 $ \\
SPT-CL J0423-5506 & $   65.8153 $  & $  -55.1036 $  & $ 68 \pm 22 $ & $   4.06 $  & $   4.42 $  & $   3.91 $  & $   4.51 $  & $ 1.25 $  & $ 0.20 \pm 0.03  $ & $   2.71 \pm   0.93 $ \\
SPT-CL J0423-6143 & $   65.9366 $  & $  -61.7183 $  & $ 76 \pm 20 $ & $   4.46 $  & $   4.03 $  & $   3.12 $  & $   4.65 $  & $ 0.25 $  & $ 0.71 \pm 0.04  $ & $   2.46 \pm   0.82 $ \\
SPT-CL J0426-5455 & $   66.5205 $  & $  -54.9201 $  & $ 163 \pm 19 $ & $   8.86 $  & $   7.94 $  & $   6.70 $  & $   8.86 $  & $ 0.50 $  & $ 0.62 \pm 0.04  $ & $   5.00 \pm   1.00 $ \\
SPT-CL J0428-6049 & $   67.0291 $  & $  -60.8302 $  & $ 89 \pm 21 $ & $   4.74 $  & $   4.89 $  & $   4.20 $  & $   5.06 $  & $ 1.25 $  & $> 1.11^A  $ & -\\
SPT-CL J0430-6251 & $   67.7086 $  & $  -62.8536 $  & $ 86 \pm 19 $ & $   5.16 $  & $   4.56 $  & $   3.48 $  & $   5.20 $  & $ 0.25 $  & $ 0.38 \pm 0.05  $ & $   3.14 \pm   0.86 $ \\
SPT-CL J0431-6126 & $   67.8393 $  & $  -61.4438 $  & $ 321 \pm 54 $ & $   4.24 $  & $   5.48 $  & $   6.36 $  & $   6.40 $  & $ 3.00 $  & $ 0.0577 $ & $   4.14 \pm   1.00 $ \\
SPT-CL J0433-5630 & $   68.2522 $  & $  -56.5038 $  & $ 102 \pm 21 $ & $   5.02 $  & $   5.34 $  & $   5.27 $  & $   5.35 $  & $ 1.75 $  & $ 0.6920 $ & $   2.96 \pm   0.82 $ \\
SPT-CL J0441-5859 & $   70.4411 $  & $  -58.9931 $  & $ 88 \pm 22 $ & $   3.89 $  & $   4.36 $  & $   4.52 $  & $   4.54 $  & $ 2.25 $  & $> 1.06^A  $ & -\\
SPT-CL J0444-5603 & $   71.1130 $  & $  -56.0566 $  & $ 88 \pm 19 $ & $   5.19 $  & $   4.21 $  & $   3.12 $  & $   5.30 $  & $ 0.25 $  & $ 0.98 \pm 0.10  $ & $   2.71 \pm   0.79 $ \\
SPT-CL J0446-5849 & $   71.5160 $  & $  -58.8226 $  & $ 136 \pm 19 $ & $   7.34 $  & $   6.29 $  & $   4.95 $  & $   7.44 $  & $ 0.25 $  & $ 1.16 \pm 0.11  $ & $   3.75 \pm   0.82 $ \\
SPT-CL J0452-5945 & $   73.1282 $  & $  -59.7622 $  & $ 85 \pm 23 $ & $   3.78 $  & $   4.37 $  & $   4.50 $  & $   4.50 $  & $ 2.50 $  & $> 0.66^A  $ & -\\
SPT-CL J0456-5623 & $   74.1745 $  & $  -56.3869 $  & $ 79 \pm 19 $ & $   4.76 $  & $   4.32 $  & $   3.41 $  & $   4.76 $  & $ 0.50 $  & $ 0.64 \pm 0.04  $ & $   2.61 \pm   0.82 $ \\
SPT-CL J0456-6141 & $   74.1496 $  & $  -61.6840 $  & $ 80 \pm 19 $ & $   4.79 $  & $   4.05 $  & $   3.42 $  & $   4.84 $  & $ 0.25 $  & $ 0.39 \pm 0.03  $ & $   2.86 \pm   0.86 $ \\
SPT-CL J0458-5741 & $   74.6021 $  & $  -57.6952 $  & $ 85 \pm 23 $ & $   3.94 $  & $   4.56 $  & $   4.91 $  & $   4.91 $  & $ 2.50 $  & $> 1.03^A  $ & -\\
SPT-CL J0502-6113 & $   75.5400 $  & $  -61.2315 $  & $ 79 \pm 19 $ & $   5.02 $  & $   4.41 $  & $   3.54 $  & $   5.09 $  & $ 0.25 $  & $ 0.67 \pm 0.05  $ & $   2.82 \pm   0.82 $ \\
SPT-CL J0509-5342* & $   77.3360 $  & $  -53.7045 $  & $ 157 \pm 25 $ & $   6.61 $  & $   6.04 $  & $   5.09 $  & $   6.61 $  & $ 0.50 $  & $ 0.4626 $ & $   5.36 \pm   0.71 $ \\
SPT-CL J0511-5154 & $   77.9202 $  & $  -51.9044 $  & $ 119 \pm 24 $ & $   5.63 $  & $   4.73 $  & $   3.86 $  & $   5.63 $  & $ 0.50 $  & $ 0.6450 $ & $   3.71 \pm   0.93 $ \\
SPT-CL J0514-5118 & $   78.6859 $  & $  -51.3100 $  & $ 111 \pm 29 $ & $   4.61 $  & $   4.82 $  & $   4.52 $  & $   4.82 $  & $ 1.50 $  & $> 1.16^A  $ & -\\
SPT-CL J0516-5430 & $   79.1480 $  & $  -54.5062 $  & $ 241 \pm 26 $ & $   9.11 $  & $   9.37 $  & $   8.57 $  & $   9.42 $  & $ 0.75 $  & $ 0.2950 $ & $   6.57 \pm   1.36 $ \\
SPT-CL J0521-5104 & $   80.2983 $  & $  -51.0812 $  & $ 134 \pm 28 $ & $   5.34 $  & $   5.28 $  & $   4.96 $  & $   5.45 $  & $ 1.00 $  & $ 0.6755 $ & $   3.54 \pm   0.96 $ \\
SPT-CL J0522-5026 & $   80.5190 $  & $  -50.4409 $  & $ 121 \pm 32 $ & $   4.50 $  & $   4.82 $  & $   4.72 $  & $   4.87 $  & $ 1.75 $  & $ 0.53 \pm 0.04  $ & $   3.21 \pm   1.00 $ \\
SPT-CL J0527-5928 & $   81.8111 $  & $  -59.4833 $  & $ 80 \pm 25 $ & $   4.52 $  & $   3.66 $  & $   3.14 $  & $   4.71 $  & $ 0.25 $  & $> 0.93^A  $ & -\\
SPT-CL J0528-5300* & $   82.0173 $  & $  -53.0001 $  & $ 110 \pm 23 $ & $   5.42 $  & $   4.38 $  & $   3.52 $  & $   5.45 $  & $ 0.25 $  & $ 0.7648 $ & $   3.21 \pm   0.57 $ \\
SPT-CL J0529-5238 & $   82.2923 $  & $  -52.6417 $  & $ 86 \pm 24 $ & $   4.31 $  & $   3.50 $  & $   2.68 $  & $   4.52 $  & $ 0.25 $  & $> 1.16^A  $ & -\\
SPT-CL J0532-5647 & $   83.1586 $  & $  -56.7893 $  & $ 99 \pm 32 $ & $   3.19 $  & $   4.09 $  & $   4.41 $  & $   4.51 $  & $ 2.75 $  & $> 0.93^A  $ & -\\
SPT-CL J0533-5005* & $   83.3984 $  & $  -50.0918 $  & $ 116 \pm 24 $ & $   5.51 $  & $   5.08 $  & $   4.32 $  & $   5.59 $  & $ 0.25 $  & $ 0.8810 $ & $   2.75 \pm   0.61 $ \\
SPT-CL J0534-5937 & $   83.6018 $  & $  -59.6289 $  & $ 82 \pm 25 $ & $   4.25 $  & $   3.53 $  & $   2.78 $  & $   4.57 $  & $ 0.25 $  & $ 0.5761 $ & $   2.86 \pm   1.00 $ \\
SPT-CL J0537-5549 & $   84.2578 $  & $  -55.8268 $  & $ 100 \pm 30 $ & $   3.91 $  & $   4.51 $  & $   4.55 $  & $   4.55 $  & $ 2.00 $  & $> 1.11^A  $ & -\\
SPT-CL J0538-5657 & $   84.5865 $  & $  -56.9530 $  & $ 102 \pm 30 $ & $   3.79 $  & $   4.37 $  & $   4.61 $  & $   4.63 $  & $ 2.75 $  & $> 1.50^A  $ & -\\
SPT-CL J0539-5744 & $   84.9998 $  & $  -57.7432 $  & $ 109 \pm 25 $ & $   5.01 $  & $   4.61 $  & $   3.86 $  & $   5.12 $  & $ 0.25 $  & $ 0.76 \pm 0.03  $ & $   3.18 \pm   0.96 $ \\
SPT-CL J0546-5345* & $   86.6541 $  & $  -53.7615 $  & $ 173 \pm 24 $ & $   7.69 $  & $   6.99 $  & $   6.20 $  & $   7.69 $  & $ 0.50 $  & $ 1.0670 $ & $   5.29 \pm   0.71 $ \\
SPT-CL J0551-5709* & $   87.9016 $  & $  -57.1565 $  & $ 150 \pm 28 $ & $   6.00 $  & $   6.06 $  & $   5.48 $  & $   6.13 $  & $ 1.00 $  & $ 0.4230 $ & $   3.82 \pm   0.54 $ \\
SPT-CL J0556-5403 & $   89.2016 $  & $  -54.0630 $  & $ 100 \pm 25 $ & $   4.72 $  & $   4.41 $  & $   3.79 $  & $   4.83 $  & $ 0.25 $  & $ 0.93 \pm 0.06  $ & $   2.79 \pm   0.93 $ \\
SPT-CL J0559-5249* & $   89.9245 $  & $  -52.8265 $  & $ 228 \pm 26 $ & $   8.81 $  & $   9.15 $  & $   8.51 $  & $   9.28 $  & $ 1.00 $  & $ 0.6112 $ & $   6.79 \pm   0.86 $ \\
SPT-CL J2002-5335 & $  300.5113 $  & $  -53.5913 $  & $ 75 \pm 23 $ & $   4.44 $  & $   4.30 $  & $   4.01 $  & $   4.53 $  & $ 1.25 $  & $> 1.02^A  $ & -\\
SPT-CL J2005-5635 & $  301.3385 $  & $  -56.5902 $  & $ 80 \pm 19 $ & $   4.68 $  & $   4.38 $  & $   3.88 $  & $   4.68 $  & $ 0.50 $  & $> 0.64^A  $ & -\\
SPT-CL J2006-5325 & $  301.6620 $  & $  -53.4287 $  & $ 86 \pm 23 $ & $   4.93 $  & $   4.77 $  & $   4.01 $  & $   5.06 $  & $ 1.00 $  & $> 1.50^A  $ & -\\
SPT-CL J2007-4906 & $  301.9663 $  & $  -49.1105 $  & $ 87 \pm 23 $ & $   4.46 $  & $   3.84 $  & $   3.14 $  & $   4.50 $  & $ 0.25 $  & $ 1.25 \pm 0.11  $ & $   2.11 \pm   0.82 $ \\
SPT-CL J2009-5756 & $  302.4262 $  & $  -57.9480 $  & $ 80 \pm 19 $ & $   4.68 $  & $   4.20 $  & $   3.42 $  & $   4.68 $  & $ 0.50 $  & $ 0.63 \pm 0.03  $ & $   2.46 \pm   0.82 $ \\
SPT-CL J2011-5228 & $  302.7810 $  & $  -52.4734 $  & $ 75 \pm 23 $ & $   4.46 $  & $   4.21 $  & $   3.70 $  & $   4.55 $  & $ 0.75 $  & $ 0.98 \pm 0.06  $ & $   2.36 \pm   0.86 $ \\
SPT-CL J2011-5725 & $  302.8526 $  & $  -57.4214 $  & $ 91 \pm 19 $ & $   5.35 $  & $   5.25 $  & $   4.83 $  & $   5.43 $  & $ 0.75 $  & $ 0.2786 $ & $   3.29 \pm   0.86 $ \\
SPT-CL J2012-5342 & $  303.0822 $  & $  -53.7137 $  & $ 73 \pm 23 $ & $   4.65 $  & $   4.38 $  & $   3.84 $  & $   4.65 $  & $ 0.50 $  & $> 0.68^A  $ & -\\
SPT-CL J2012-5649 & $  303.1132 $  & $  -56.8308 $  & $ 116 \pm 25 $ & $   4.70 $  & $   5.83 $  & $   5.99 $  & $   5.99 $  & $ 2.50 $  & $ 0.0552 $ & $   3.79 \pm   0.93 $ \\
SPT-CL J2013-5432 & $  303.4968 $  & $  -54.5445 $  & $ 78 \pm 23 $ & $   4.23 $  & $   3.05 $  & $   1.92 $  & $   4.75 $  & $ 0.25 $  & $> 1.02^A  $ & -\\
SPT-CL J2015-5504 & $  303.9884 $  & $  -55.0715 $  & $ 79 \pm 23 $ & $   4.64 $  & $   4.28 $  & $   3.68 $  & $   4.64 $  & $ 0.50 $  & $> 0.61^A  $ & -\\
SPT-CL J2016-4954 & $  304.0181 $  & $  -49.9122 $  & $ 100 \pm 23 $ & $   5.01 $  & $   4.62 $  & $   3.77 $  & $   5.01 $  & $ 0.50 $  & $ 0.26 \pm 0.03  $ & $   3.36 \pm   1.00 $ \\
SPT-CL J2017-6258 & $  304.4827 $  & $  -62.9763 $  & $ 117 \pm 22 $ & $   5.89 $  & $   6.45 $  & $   5.92 $  & $   6.45 $  & $ 1.50 $  & $ 0.54 \pm 0.04  $ & $   3.71 \pm   0.86 $ \\
SPT-CL J2018-4528 & $  304.6076 $  & $  -45.4807 $  & $ 85 \pm 23 $ & $   4.59 $  & $   4.26 $  & $   3.41 $  & $   4.64 $  & $ 0.25 $  & $ 0.41 \pm 0.03  $ & $   2.93 \pm   0.93 $ \\
SPT-CL J2019-5642 & $  304.7703 $  & $  -56.7079 $  & $ 94 \pm 19 $ & $   5.17 $  & $   5.05 $  & $   4.41 $  & $   5.25 $  & $ 0.75 $  & $ 0.14 \pm 0.03  $ & $   3.25 \pm   0.89 $ \\
SPT-CL J2020-4646 & $  305.1936 $  & $  -46.7702 $  & $ 97 \pm 24 $ & $   5.07 $  & $   5.08 $  & $   4.38 $  & $   5.09 $  & $ 1.25 $  & $ 0.19 \pm 0.02  $ & $   3.50 \pm   1.00 $ \\
SPT-CL J2020-6314 & $  305.0301 $  & $  -63.2413 $  & $ 82 \pm 19 $ & $   5.31 $  & $   4.69 $  & $   3.84 $  & $   5.37 $  & $ 0.25 $  & $ 0.55 \pm 0.03  $ & $   3.04 \pm   0.82 $ \\
SPT-CL J2021-5256 & $  305.4690 $  & $  -52.9439 $  & $ 190 \pm 54 $ & $   3.44 $  & $   4.58 $  & $   5.27 $  & $   5.31 $  & $ 3.00 $  & $ 0.11 \pm 0.02  $ & $   3.71 \pm   1.00 $ \\
SPT-CL J2022-6323 & $  305.5235 $  & $  -63.3973 $  & $ 106 \pm 19 $ & $   6.58 $  & $   5.91 $  & $   5.04 $  & $   6.58 $  & $ 0.50 $  & $ 0.3830 $ & $   3.89 \pm   0.89 $ \\
SPT-CL J2023-5535 & $  305.8377 $  & $  -55.5903 $  & $ 282 \pm 24 $ & $  11.75 $  & $  13.36 $  & $  13.04 $  & $  13.41 $  & $ 1.75 $  & $ 0.2320 $ & $   7.25 \pm   1.39 $ \\
SPT-CL J2025-5117 & $  306.4837 $  & $  -51.2904 $  & $ 183 \pm 22 $ & $   9.37 $  & $   8.81 $  & $   7.36 $  & $   9.48 $  & $ 0.75 $  & $ 0.18 \pm 0.02  $ & $   6.36 \pm   1.29 $ \\
SPT-CL J2026-4513 & $  306.6140 $  & $  -45.2256 $  & $ 107 \pm 22 $ & $   5.53 $  & $   5.09 $  & $   4.26 $  & $   5.53 $  & $ 0.50 $  & $ 0.74 \pm 0.03  $ & $   3.36 \pm   0.86 $ \\
SPT-CL J2030-5638 & $  307.7067 $  & $  -56.6352 $  & $ 100 \pm 20 $ & $   5.28 $  & $   5.33 $  & $   4.89 $  & $   5.47 $  & $ 1.00 $  & $ 0.40 \pm 0.04  $ & $   3.21 \pm   0.86 $ \\
SPT-CL J2032-5627 & $  308.0800 $  & $  -56.4557 $  & $ 167 \pm 22 $ & $   7.64 $  & $   8.03 $  & $   7.92 $  & $   8.14 $  & $ 1.75 $  & $ 0.2840 $ & $   4.82 \pm   1.04 $ \\
SPT-CL J2034-5936 & $  308.5408 $  & $  -59.6007 $  & $ 144 \pm 18 $ & $   8.54 $  & $   7.62 $  & $   6.23 $  & $   8.57 $  & $ 0.25 $  & $ 0.92 \pm 0.10  $ & $   4.36 \pm   0.86 $ \\
SPT-CL J2035-5251 & $  308.8026 $  & $  -52.8527 $  & $ 205 \pm 23 $ & $   9.60 $  & $   9.95 $  & $   8.97 $  & $  10.00 $  & $ 0.75 $  & $ 0.47 \pm 0.03  $ & $   6.25 \pm   1.25 $ \\
SPT-CL J2035-5614 & $  308.9022 $  & $  -56.2407 $  & $ 76 \pm 19 $ & $   4.43 $  & $   4.27 $  & $   3.70 $  & $   4.55 $  & $ 0.75 $  & $> 1.02^A  $ & -\\
SPT-CL J2039-5723 & $  309.8246 $  & $  -57.3871 $  & $ 76 \pm 19 $ & $   4.69 $  & $   4.34 $  & $   4.13 $  & $   4.69 $  & $ 0.50 $  & $> 1.23^A  $ & -\\
SPT-CL J2040-4451 & $  310.2468 $  & $  -44.8599 $  & $ 122 \pm 22 $ & $   6.08 $  & $   4.60 $  & $   3.62 $  & $   6.28 $  & $ 0.25 $  & $ 1.37 \pm 0.12  $ & $   3.21 \pm   0.79 $ \\
SPT-CL J2040-5230 & $  310.1255 $  & $  -52.5052 $  & $ 76 \pm 22 $ & $   4.50 $  & $   3.41 $  & $   2.57 $  & $   4.70 $  & $ 0.25 $  & $> 1.01^A  $ & -\\
SPT-CL J2040-5342 & $  310.2195 $  & $  -53.7122 $  & $ 107 \pm 22 $ & $   5.88 $  & $   5.36 $  & $   4.61 $  & $   5.88 $  & $ 0.50 $  & $ 0.56 \pm 0.04  $ & $   3.79 \pm   0.93 $ \\
SPT-CL J2040-5725 & $  310.0631 $  & $  -57.4287 $  & $ 107 \pm 19 $ & $   6.38 $  & $   5.91 $  & $   5.03 $  & $   6.38 $  & $ 0.50 $  & $ 0.9300 $ & $   3.29 \pm   0.79 $ \\
SPT-CL J2043-5035 & $  310.8285 $  & $  -50.5929 $  & $ 151 \pm 22 $ & $   7.81 $  & $   7.47 $  & $   6.73 $  & $   7.81 $  & $ 0.50 $  & $ 0.7234 $ & $   4.79 \pm   1.00 $ \\
SPT-CL J2043-5614 & $  310.7906 $  & $  -56.2351 $  & $ 74 \pm 19 $ & $   4.66 $  & $   4.00 $  & $   3.17 $  & $   4.72 $  & $ 0.25 $  & $ 0.66 \pm 0.05  $ & $   2.50 \pm   0.79 $ \\
SPT-CL J2045-6026 & $  311.3649 $  & $  -60.4469 $  & $ 67 \pm 19 $ & $   4.59 $  & $   3.46 $  & $   2.51 $  & $   4.77 $  & $ 0.25 $  & $> 0.47^A  $ & -\\
SPT-CL J2046-4542 & $  311.5620 $  & $  -45.7111 $  & $ 92 \pm 23 $ & $   4.43 $  & $   4.36 $  & $   3.73 $  & $   4.54 $  & $ 1.00 $  & $> 1.02^A  $ & -\\
SPT-CL J2048-4524 & $  312.2268 $  & $  -45.4150 $  & $ 89 \pm 24 $ & $   4.25 $  & $   4.44 $  & $   3.96 $  & $   4.56 $  & $ 1.00 $  & $> 0.98^A  $ & -\\
SPT-CL J2051-6256 & $  312.8027 $  & $  -62.9348 $  & $ 83 \pm 20 $ & $   4.97 $  & $   5.04 $  & $   4.39 $  & $   5.17 $  & $ 1.25 $  & $ 0.47 \pm 0.03  $ & $   2.96 \pm   0.82 $ \\
SPT-CL J2055-5456 & $  313.9953 $  & $  -54.9369 $  & $ 113 \pm 18 $ & $   6.52 $  & $   6.12 $  & $   5.20 $  & $   6.52 $  & $ 0.50 $  & $ 0.13 \pm 0.02  $ & $   4.07 \pm   0.93 $ \\
SPT-CL J2056-5106 & $  314.0723 $  & $  -51.1163 $  & $ 89 \pm 25 $ & $   4.37 $  & $   4.60 $  & $   4.63 $  & $   4.70 $  & $ 2.00 $  & $> 1.02^A  $ & -\\
SPT-CL J2056-5459 & $  314.2186 $  & $  -54.9933 $  & $ 102 \pm 19 $ & $   5.95 $  & $   5.47 $  & $   4.89 $  & $   5.95 $  & $ 0.50 $  & $ 0.7180 $ & $   3.21 \pm   0.79 $ \\
SPT-CL J2057-5251 & $  314.4105 $  & $  -52.8567 $  & $ 69 \pm 23 $ & $   4.50 $  & $   4.03 $  & $   3.38 $  & $   4.52 $  & $ 0.25 $  & $> 1.50^A  $ & -\\
SPT-CL J2058-5608 & $  314.5893 $  & $  -56.1454 $  & $ 78 \pm 18 $ & $   4.84 $  & $   4.03 $  & $   2.88 $  & $   5.02 $  & $ 0.25 $  & $ 0.6060 $ & $   2.71 \pm   0.79 $ \\
SPT-CL J2059-5018 & $  314.9324 $  & $  -50.3049 $  & $ 91 \pm 22 $ & $   4.73 $  & $   4.21 $  & $   3.30 $  & $   4.79 $  & $ 0.25 $  & $ 0.41 \pm 0.03  $ & $   3.04 \pm   0.96 $ \\
SPT-CL J2100-4548 & $  315.0936 $  & $  -45.8057 $  & $ 90 \pm 23 $ & $   4.84 $  & $   4.53 $  & $   4.12 $  & $   4.84 $  & $ 0.50 $  & $ 0.7121 $ & $   2.82 \pm   0.89 $ \\
SPT-CL J2100-5708 & $  315.1503 $  & $  -57.1347 $  & $ 83 \pm 19 $ & $   5.03 $  & $   4.47 $  & $   3.21 $  & $   5.11 $  & $ 0.25 $  & $ 0.55 \pm 0.03  $ & $   2.82 \pm   0.82 $ \\
SPT-CL J2101-5542 & $  315.3106 $  & $  -55.7027 $  & $ 115 \pm 29 $ & $   4.67 $  & $   4.99 $  & $   4.82 $  & $   5.04 $  & $ 1.75 $  & $ 0.20 \pm 0.02  $ & $   3.04 \pm   0.89 $ \\
SPT-CL J2101-6123 & $  315.4594 $  & $  -61.3972 $  & $ 84 \pm 19 $ & $   5.28 $  & $   4.83 $  & $   3.95 $  & $   5.28 $  & $ 0.50 $  & $ 0.59 \pm 0.04  $ & $   2.93 \pm   0.79 $ \\
SPT-CL J2103-5411 & $  315.7687 $  & $  -54.1951 $  & $ 78 \pm 22 $ & $   4.72 $  & $   4.32 $  & $   3.48 $  & $   4.88 $  & $ 0.25 $  & $ 0.49 \pm 0.03  $ & $   3.07 \pm   0.93 $ \\
SPT-CL J2104-5224 & $  316.2283 $  & $  -52.4044 $  & $ 77 \pm 23 $ & $   4.97 $  & $   3.54 $  & $   2.19 $  & $   5.32 $  & $ 0.25 $  & $ 0.7991 $ & $   3.14 \pm   0.86 $ \\
SPT-CL J2106-5820 & $  316.5144 $  & $  -58.3459 $  & $ 66 \pm 19 $ & $   4.43 $  & $   2.94 $  & $   1.56 $  & $   4.81 $  & $ 0.25 $  & $> 1.00^A  $ & -\\
SPT-CL J2106-5844 & $  316.5210 $  & $  -58.7448 $  & $ 363 \pm 18 $ & $  21.78 $  & $  18.67 $  & $  14.36 $  & $  22.08 $  & $ 0.25 $  & $ 1.1320 $ & $   8.39 \pm   1.68 $ \\
SPT-CL J2106-6019 & $  316.6642 $  & $  -60.3299 $  & $ 73 \pm 19 $ & $   4.82 $  & $   3.49 $  & $   2.21 $  & $   4.98 $  & $ 0.25 $  & $ 0.97 \pm 0.05  $ & $   2.39 \pm   0.75 $ \\
SPT-CL J2106-6303 & $  316.6596 $  & $  -63.0510 $  & $ 84 \pm 20 $ & $   4.56 $  & $   4.82 $  & $   4.24 $  & $   4.90 $  & $ 1.25 $  & $> 1.04^A  $ & -\\
SPT-CL J2109-4626 & $  317.4516 $  & $  -46.4370 $  & $ 104 \pm 21 $ & $   5.32 $  & $   4.33 $  & $   3.24 $  & $   5.51 $  & $ 0.25 $  & $ 0.98 \pm 0.10  $ & $   3.11 \pm   0.82 $ \\
SPT-CL J2109-5040 & $  317.3820 $  & $  -50.6773 $  & $ 103 \pm 27 $ & $   4.56 $  & $   5.14 $  & $   5.06 $  & $   5.17 $  & $ 2.00 $  & $ 0.49 \pm 0.04  $ & $   3.29 \pm   0.93 $ \\
SPT-CL J2110-5244 & $  317.5502 $  & $  -52.7486 $  & $ 114 \pm 22 $ & $   6.22 $  & $   5.82 $  & $   4.79 $  & $   6.22 $  & $ 0.50 $  & $ 0.63 \pm 0.03  $ & $   3.93 \pm   0.93 $ \\
SPT-CL J2111-5338 & $  317.9216 $  & $  -53.6496 $  & $ 111 \pm 25 $ & $   5.53 $  & $   5.51 $  & $   4.80 $  & $   5.65 $  & $ 1.00 $  & $ 0.44 \pm 0.04  $ & $   3.71 \pm   0.93 $ \\
SPT-CL J2115-4659 & $  318.7995 $  & $  -46.9862 $  & $ 140 \pm 34 $ & $   4.98 $  & $   5.40 $  & $   5.48 $  & $   5.60 $  & $ 2.25 $  & $ 0.35 \pm 0.03  $ & $   3.75 \pm   0.96 $ \\
SPT-CL J2118-5055 & $  319.7291 $  & $  -50.9329 $  & $ 116 \pm 22 $ & $   5.52 $  & $   4.84 $  & $   3.66 $  & $   5.62 $  & $ 0.25 $  & $ 0.6254 $ & $   3.54 \pm   0.89 $ \\
SPT-CL J2119-6230 & $  319.8846 $  & $  -62.5096 $  & $ 61 \pm 19 $ & $   4.37 $  & $   3.35 $  & $   2.83 $  & $   4.55 $  & $ 0.25 $  & $ 0.69 \pm 0.04  $ & $   2.36 \pm   0.79 $ \\
SPT-CL J2120-4728 & $  320.1594 $  & $  -47.4776 $  & $ 109 \pm 21 $ & $   5.87 $  & $   4.82 $  & $   3.61 $  & $   5.98 $  & $ 0.25 $  & $ 0.99 \pm 0.10  $ & $   3.43 \pm   0.86 $ \\
SPT-CL J2121-5546 & $  320.2715 $  & $  -55.7780 $  & $ 85 \pm 19 $ & $   4.61 $  & $   4.44 $  & $   3.55 $  & $   4.79 $  & $ 0.75 $  & $> 0.75^A  $ & -\\
SPT-CL J2121-6335 & $  320.4269 $  & $  -63.5843 $  & $ 133 \pm 32 $ & $   3.87 $  & $   5.01 $  & $   5.43 $  & $   5.43 $  & $ 2.75 $  & $ 0.23 \pm 0.03  $ & $   3.29 \pm   0.86 $ \\
SPT-CL J2124-6124 & $  321.1488 $  & $  -61.4141 $  & $ 147 \pm 21 $ & $   8.08 $  & $   8.18 $  & $   7.70 $  & $   8.21 $  & $ 1.00 $  & $ 0.4350 $ & $   4.71 \pm   1.00 $ \\
SPT-CL J2125-6113 & $  321.2902 $  & $  -61.2292 $  & $ 76 \pm 19 $ & $   4.74 $  & $   4.55 $  & $   4.27 $  & $   4.74 $  & $ 0.50 $  & $> 1.50^A  $ & -\\
SPT-CL J2127-6443 & $  321.9939 $  & $  -64.7288 $  & $ 75 \pm 22 $ & $   3.81 $  & $   4.44 $  & $   4.53 $  & $   4.54 $  & $ 1.75 $  & $> 0.97^A  $ & -\\
SPT-CL J2130-4737 & $  322.6622 $  & $  -47.6257 $  & $ 81 \pm 23 $ & $   4.51 $  & $   3.37 $  & $   2.18 $  & $   4.83 $  & $ 0.25 $  & $> 1.50^A  $ & -\\
SPT-CL J2130-6458 & $  322.7285 $  & $  -64.9764 $  & $ 130 \pm 20 $ & $   7.31 $  & $   7.31 $  & $   6.43 $  & $   7.57 $  & $ 1.00 $  & $ 0.3160 $ & $   4.54 \pm   0.96 $ \\
SPT-CL J2131-5003 & $  322.9717 $  & $  -50.0647 $  & $ 88 \pm 23 $ & $   4.83 $  & $   4.50 $  & $   3.98 $  & $   4.83 $  & $ 0.50 $  & $ 0.46 \pm 0.03  $ & $   3.04 \pm   0.96 $ \\
SPT-CL J2133-5411 & $  323.2978 $  & $  -54.1845 $  & $ 71 \pm 22 $ & $   4.17 $  & $   3.00 $  & $   1.75 $  & $   4.58 $  & $ 0.25 $  & $> 1.50^A  $ & -\\
SPT-CL J2135-5452 & $  323.9060 $  & $  -54.8773 $  & $ 90 \pm 20 $ & $   4.39 $  & $   4.57 $  & $   4.00 $  & $   4.61 $  & $ 1.00 $  & $> 1.00^A  $ & -\\
SPT-CL J2135-5726 & $  323.9158 $  & $  -57.4415 $  & $ 176 \pm 18 $ & $  10.43 $  & $   9.81 $  & $   8.64 $  & $  10.43 $  & $ 0.50 $  & $ 0.4270 $ & $   5.75 \pm   1.11 $ \\
SPT-CL J2136-4704 & $  324.1175 $  & $  -47.0803 $  & $ 114 \pm 22 $ & $   6.10 $  & $   5.46 $  & $   4.68 $  & $   6.17 $  & $ 0.25 $  & $ 0.4250 $ & $   4.11 \pm   0.96 $ \\
SPT-CL J2136-5519 & $  324.2392 $  & $  -55.3215 $  & $ 73 \pm 19 $ & $   4.65 $  & $   3.91 $  & $   3.37 $  & $   4.65 $  & $ 0.50 $  & $> 1.50^A  $ & -\\
SPT-CL J2136-5535 & $  324.0898 $  & $  -55.5853 $  & $ 74 \pm 19 $ & $   4.52 $  & $   4.40 $  & $   3.79 $  & $   4.58 $  & $ 0.75 $  & $> 1.19^A  $ & -\\
SPT-CL J2136-5723 & $  324.1209 $  & $  -57.3923 $  & $ 75 \pm 20 $ & $   4.22 $  & $   4.55 $  & $   4.10 $  & $   4.55 $  & $ 1.50 $  & $> 1.04^A  $ & -\\
SPT-CL J2136-6307 & $  324.2334 $  & $  -63.1233 $  & $ 100 \pm 19 $ & $   6.11 $  & $   5.80 $  & $   4.91 $  & $   6.25 $  & $ 0.75 $  & $ 0.9260 $ & $   3.25 \pm   0.75 $ \\
SPT-CL J2137-6437 & $  324.4178 $  & $  -64.6235 $  & $ 71 \pm 19 $ & $   4.40 $  & $   4.10 $  & $   3.27 $  & $   4.60 $  & $ 0.75 $  & $ 0.91 \pm 0.10  $ & $   2.18 \pm   0.75 $ \\
SPT-CL J2138-6007 & $  324.5060 $  & $  -60.1324 $  & $ 225 \pm 19 $ & $  12.39 $  & $  12.41 $  & $  11.14 $  & $  12.64 $  & $ 0.75 $  & $ 0.3190 $ & $   6.82 \pm   1.32 $ \\
SPT-CL J2139-5420 & $  324.9670 $  & $  -54.3396 $  & $ 77 \pm 24 $ & $   4.69 $  & $   4.69 $  & $   3.92 $  & $   4.81 $  & $ 0.75 $  & $ 0.23 \pm 0.03  $ & $   3.21 \pm   1.00 $ \\
SPT-CL J2140-5331 & $  325.0304 $  & $  -53.5199 $  & $ 90 \pm 23 $ & $   4.54 $  & $   4.13 $  & $   3.48 $  & $   4.55 $  & $ 0.25 $  & $ 0.53 \pm 0.03  $ & $   2.75 \pm   0.96 $ \\
SPT-CL J2140-5727 & $  325.1380 $  & $  -57.4564 $  & $ 76 \pm 19 $ & $   4.90 $  & $   4.01 $  & $   3.14 $  & $   5.08 $  & $ 0.25 $  & $ 0.42 \pm 0.03  $ & $   2.93 \pm   0.86 $ \\
SPT-CL J2142-4846 & $  325.5693 $  & $  -48.7743 $  & $ 70 \pm 24 $ & $   4.02 $  & $   4.53 $  & $   4.36 $  & $   4.53 $  & $ 1.50 $  & $> 0.80^A  $ & -\\
SPT-CL J2145-5644 & $  326.4694 $  & $  -56.7477 $  & $ 213 \pm 18 $ & $  12.30 $  & $  11.67 $  & $  10.39 $  & $  12.30 $  & $ 0.50 $  & $ 0.4800 $ & $   6.46 \pm   1.25 $ \\
SPT-CL J2146-4633 & $  326.6473 $  & $  -46.5505 $  & $ 202 \pm 22 $ & $   9.59 $  & $   8.67 $  & $   6.99 $  & $   9.59 $  & $ 0.50 $  & $ 0.9330 $ & $   5.43 \pm   1.07 $ \\
SPT-CL J2146-4846 & $  326.5346 $  & $  -48.7774 $  & $ 115 \pm 25 $ & $   5.59 $  & $   5.88 $  & $   5.34 $  & $   5.88 $  & $ 1.50 $  & $ 0.6230 $ & $   3.71 \pm   0.93 $ \\
SPT-CL J2146-5736 & $  326.6963 $  & $  -57.6138 $  & $ 98 \pm 19 $ & $   5.94 $  & $   5.46 $  & $   4.57 $  & $   5.94 $  & $ 0.50 $  & $ 0.60 \pm 0.03  $ & $   3.32 \pm   0.82 $ \\
SPT-CL J2148-4843 & $  327.0971 $  & $  -48.7287 $  & $ 77 \pm 23 $ & $   4.19 $  & $   2.88 $  & $   1.81 $  & $   4.64 $  & $ 0.25 $  & $ 0.98 \pm 0.10  $ & $   2.43 \pm   0.86 $ \\
SPT-CL J2148-6116 & $  327.1798 $  & $  -61.2791 $  & $ 124 \pm 20 $ & $   6.95 $  & $   7.22 $  & $   6.31 $  & $   7.27 $  & $ 1.25 $  & $ 0.5710 $ & $   4.11 \pm   0.89 $ \\
SPT-CL J2149-5330 & $  327.3770 $  & $  -53.5014 $  & $ 95 \pm 24 $ & $   4.79 $  & $   4.50 $  & $   4.04 $  & $   4.79 $  & $ 0.50 $  & $ 0.53 \pm 0.03  $ & $   2.93 \pm   0.93 $ \\
SPT-CL J2150-6111 & $  327.7177 $  & $  -61.1954 $  & $ 73 \pm 21 $ & $   4.12 $  & $   4.50 $  & $   4.70 $  & $   4.70 $  & $ 2.50 $  & $> 1.11^A  $ & -\\
SPT-CL J2152-4629 & $  328.1943 $  & $  -46.4947 $  & $ 94 \pm 21 $ & $   5.45 $  & $   4.56 $  & $   3.26 $  & $   5.60 $  & $ 0.25 $  & $> 1.50^A  $ & -\\
SPT-CL J2152-5143 & $  328.0034 $  & $  -51.7245 $  & $ 67 \pm 24 $ & $   4.45 $  & $   4.33 $  & $   3.79 $  & $   4.53 $  & $ 0.75 $  & $ 0.40 \pm 0.03  $ & $   2.82 \pm   0.96 $ \\
SPT-CL J2152-5633 & $  328.1458 $  & $  -56.5641 $  & $ 100 \pm 21 $ & $   5.16 $  & $   5.66 $  & $   5.68 $  & $   5.84 $  & $ 1.75 $  & $> 1.50^A  $ & -\\
SPT-CL J2155-5103 & $  328.8747 $  & $  -51.0508 $  & $ 73 \pm 25 $ & $   4.11 $  & $   4.41 $  & $   4.40 $  & $   4.52 $  & $ 1.75 $  & $> 1.06^A  $ & -\\
SPT-CL J2155-5225 & $  328.8941 $  & $  -52.4169 $  & $ 95 \pm 25 $ & $   4.45 $  & $   4.77 $  & $   4.36 $  & $   4.77 $  & $ 1.50 $  & $ 0.59 \pm 0.04  $ & $   2.86 \pm   0.93 $ \\
SPT-CL J2155-6048 & $  328.9851 $  & $  -60.8072 $  & $ 88 \pm 20 $ & $   4.87 $  & $   5.19 $  & $   4.56 $  & $   5.24 $  & $ 1.00 $  & $ 0.5390 $ & $   2.93 \pm   0.79 $ \\
SPT-CL J2158-4702 & $  329.6901 $  & $  -47.0348 $  & $ 78 \pm 23 $ & $   4.50 $  & $   4.38 $  & $   4.17 $  & $   4.56 $  & $ 1.00 $  & $> 0.90^A  $ & -\\
SPT-CL J2158-4851 & $  329.5737 $  & $  -48.8536 $  & $ 80 \pm 23 $ & $   4.28 $  & $   3.38 $  & $   2.20 $  & $   4.61 $  & $ 0.25 $  & $> 0.75^A  $ & -\\
SPT-CL J2158-5615 & $  329.5975 $  & $  -56.2588 $  & $ 88 \pm 20 $ & $   4.28 $  & $   4.42 $  & $   3.93 $  & $   4.54 $  & $ 1.25 $  & $> 1.07^A  $ & -\\
SPT-CL J2158-6319 & $  329.6390 $  & $  -63.3175 $  & $ 62 \pm 19 $ & $   4.33 $  & $   3.43 $  & $   2.64 $  & $   4.54 $  & $ 0.25 $  & $> 1.06^A  $ & -\\
SPT-CL J2159-6244 & $  329.9922 $  & $  -62.7420 $  & $ 108 \pm 21 $ & $   6.02 $  & $   5.98 $  & $   5.54 $  & $   6.08 $  & $ 1.00 $  & $ 0.42 \pm 0.02  $ & $   3.57 \pm   0.86 $ \\
SPT-CL J2200-5547 & $  330.0304 $  & $  -55.7954 $  & $ 79 \pm 21 $ & $   3.83 $  & $   4.63 $  & $   4.72 $  & $   4.80 $  & $ 2.00 $  & $> 0.98^A  $ & -\\
SPT-CL J2201-5956 & $  330.4727 $  & $  -59.9473 $  & $ 338 \pm 25 $ & $  11.61 $  & $  13.57 $  & $  13.99 $  & $  13.99 $  & $ 2.50 $  & $ 0.0972 $ & $   7.68 \pm   1.54 $ \\
SPT-CL J2202-5936 & $  330.5483 $  & $  -59.6021 $  & $ 76 \pm 19 $ & $   4.81 $  & $   4.21 $  & $   3.36 $  & $   4.89 $  & $ 0.25 $  & $ 0.46 \pm 0.03  $ & $   2.75 \pm   0.82 $ \\
SPT-CL J2259-5432 & $  344.9820 $  & $  -54.5356 $  & $ 135 \pm 38 $ & $   4.56 $  & $   4.71 $  & $   4.65 $  & $   4.78 $  & $ 2.00 $  & $ 0.44 \pm 0.04  $ & $   3.18 \pm   1.04 $ \\
SPT-CL J2259-5617 & $  344.9974 $  & $  -56.2877 $  & $ 99 \pm 24 $ & $   5.04 $  & $   4.27 $  & $   3.55 $  & $   5.29 $  & $ 0.25 $  & $ 0.17 \pm 0.02  $ & $   3.86 \pm   1.07 $ \\
SPT-CL J2300-5331 & $  345.1765 $  & $  -53.5170 $  & $ 119 \pm 27 $ & $   5.24 $  & $   5.02 $  & $   4.65 $  & $   5.29 $  & $ 0.25 $  & $ 0.2620 $ & $   3.79 \pm   1.00 $ \\
SPT-CL J2301-5046 & $  345.4585 $  & $  -50.7823 $  & $ 92 \pm 24 $ & $   4.58 $  & $   3.83 $  & $   2.82 $  & $   4.58 $  & $ 0.50 $  & $> 1.50^A  $ & -\\
SPT-CL J2301-5546 & $  345.4688 $  & $  -55.7758 $  & $ 106 \pm 25 $ & $   5.19 $  & $   4.93 $  & $   4.62 $  & $   5.19 $  & $ 0.50 $  & $ 0.7480 $ & $   3.21 \pm   0.93 $ \\
SPT-CL J2302-5225 & $  345.6464 $  & $  -52.4329 $  & $ 104 \pm 29 $ & $   3.77 $  & $   4.24 $  & $   4.60 $  & $   4.60 $  & $ 2.50 $  & $> 1.04^A  $ & -\\
SPT-CL J2311-5011 & $  347.8427 $  & $  -50.1838 $  & $ 91 \pm 29 $ & $   3.40 $  & $   3.85 $  & $   4.42 $  & $   4.64 $  & $ 3.00 $  & $> 1.50^A  $ & -\\
SPT-CL J2312-5820 & $  348.0002 $  & $  -58.3419 $  & $ 89 \pm 24 $ & $   4.66 $  & $   3.75 $  & $   3.11 $  & $   4.78 $  & $ 0.25 $  & $ 0.88 \pm 0.09  $ & $   2.79 \pm   0.93 $ \\
SPT-CL J2329-5831 & $  352.4760 $  & $  -58.5238 $  & $ 107 \pm 25 $ & $   4.95 $  & $   4.64 $  & $   3.96 $  & $   4.95 $  & $ 0.50 $  & $ 0.82 \pm 0.04  $ & $   2.96 \pm   0.96 $ \\
SPT-CL J2331-5051* & $  352.9584 $  & $  -50.8641 $  & $ 166 \pm 23 $ & $   7.86 $  & $   6.60 $  & $   5.14 $  & $   8.04 $  & $ 0.25 $  & $ 0.5760 $ & $   5.14 \pm   0.71 $ \\
SPT-CL J2332-5358* & $  353.1040 $  & $  -53.9733 $  & $ 193 \pm 31 $ & $   7.25 $  & $   7.30 $  & $   6.84 $  & $   7.30 $  & $ 1.50 $  & $ 0.4020 $ & $   6.54 \pm   0.82 $ \\
SPT-CL J2334-5953 & $  353.6989 $  & $  -59.8892 $  & $ 94 \pm 30 $ & $   2.94 $  & $   3.98 $  & $   4.53 $  & $   4.53 $  & $ 2.50 $  & $> 1.50^A  $ & -\\
SPT-CL J2337-5942* & $  354.3544 $  & $  -59.7052 $  & $ 312 \pm 24 $ & $  14.72 $  & $  12.63 $  & $  10.11 $  & $  14.94 $  & $ 0.25 $  & $ 0.7750 $ & $   8.21 \pm   1.14 $ \\
SPT-CL J2341-5119* & $  355.2994 $  & $  -51.3328 $  & $ 227 \pm 24 $ & $   9.48 $  & $   9.02 $  & $   7.74 $  & $   9.65 $  & $ 0.75 $  & $ 1.0030 $ & $   5.61 \pm   0.82 $ \\
SPT-CL J2342-5411* & $  355.6903 $  & $  -54.1887 $  & $ 132 \pm 23 $ & $   6.18 $  & $   5.24 $  & $   3.96 $  & $   6.18 $  & $ 0.50 $  & $ 1.0750 $ & $   3.00 \pm   0.50 $ \\
SPT-CL J2343-5521 & $  355.7574 $  & $  -55.3641 $  & $ 130 \pm 28 $ & $   4.87 $  & $   5.58 $  & $   5.74 $  & $   5.74 $  & $ 2.50 $  & $> 1.50^A  $ & -\\
SPT-CL J2343-5556 & $  355.9290 $  & $  -55.9371 $  & $ 106 \pm 27 $ & $   4.49 $  & $   4.53 $  & $   4.00 $  & $   4.58 $  & $ 1.00 $  & $> 1.21^A  $ & -\\
SPT-CL J2351-5452 & $  357.8877 $  & $  -54.8753 $  & $ 151 \pm 47 $ & $   4.35 $  & $   4.70 $  & $   4.83 $  & $   4.89 $  & $ 2.75 $  & $ 0.3838 $ & $   3.32 \pm   1.04 $ \\
SPT-CL J2355-5056* & $  358.9551 $  & $  -50.9367 $  & $ 138 \pm 24 $ & $   5.73 $  & $   5.34 $  & $   4.31 $  & $   5.89 $  & $ 0.75 $  & $ 0.3196 $ & $   4.11 \pm   0.54 $ \\
SPT-CL J2359-5009* & $  359.9208 $  & $  -50.1600 $  & $ 152 \pm 27 $ & $   6.19 $  & $   6.23 $  & $   5.64 $  & $   6.35 $  & $ 1.25 $  & $ 0.7750 $ & $   3.57 \pm   0.57 $ \\